\documentclass[a4paper,fleqn,usenatbib]{mnras}
\usepackage[T1]{fontenc}
\usepackage{ae,aecompl}
\usepackage{amsmath}
\usepackage{amsfonts}
\usepackage{amssymb}
\usepackage{graphicx}
\def\mnras{MNRAS}
\def\aap{A$\&$A}
\def\nat{Nature}
\def\apj{ApJ}
\def\apjl{ApJ Letters}
\def\apjs{ApJS}

\def\icarus{Icarus}

\newcommand{\me}{\, {\rm M}_{\oplus}}
\newcommand{\msun}{\, {\rm M}_{\odot}}
\newcommand{\au}{\, {\rm au}}
\title[Proxima b formation scenarios]{Exploring plausible formation scenarios for the planet candidate orbiting Proxima Centauri}
\author[G. A. L. Coleman et al]{G. A. L. Coleman$^{1}$\thanks{Email: g.coleman@qmul.ac.uk},
R. P. Nelson$^{1}$,
S. J. Paardekooper$^{1}$,
S. Dreizler$^{2}$,
B. Giesers$^{2}$,
\newauthor G. Anglada-Escud{\'e}$^{1}$ \\
$^{1}$Astronomy Unit, Queen Mary University of London, Mile End Road, London, E1 4NS, U.K.\\
$^{2}$Institut f\"ur Astrophysik, Georg-August Universit\"at G\"ottingen, Germany\\
}
\date{}
\begin{document}
\maketitle
\begin{abstract}
We present a study of four different formation scenarios that may be able to explain the origin of the recently announced planet (`Proxima b') orbiting Proxima Centauri. The aim is to examine how the formation scenarios differ in their predictions for the multiplicity of the Proxima system, the water/volatile content of Proxima b and its orbital eccentricity, so that these can be tested by future observations. A scenario of in situ formation via giant impacts from a locally enhanced disc of planetary embryos/planetesimals predicts that Proxima b will be in a multiplanet system with a measurably finite eccentricity. Assuming that the local solid enhancement needed to form a Proxima b analogue arises because of the inwards drift of solids in the form of small planetesimals/boulders, this scenario also results in Proxima b analogues being only moderately endowed with water/volatiles. A scenario in which multiple embryos form, migrate and mutually collide within a gas disc results in Proxima b being a member of a multiple system, possibly displaying mean motion resonances, but where the constituent members are Ocean planets due to accretion occurring mainly outside of the snowline. A scenario in which a single accreting embryo forms outside the snowline, and migrates inwards while accreting planetesimals/pebbles results in Proxima b being an isolated Ocean planet on a circular orbit. A scenario in which Proxima b formed via pebble accretion interior to the snowline produces a dry planet on a circular orbit. Future observations that characterise the physical and orbital properties of Proxima b, and any additional planets in the system, will provide valuable insights into the formation history of this neighbouring planetary system.
\end{abstract}

\section{Introduction}
\label{sec:introduction}
The recent discovery of a planet (`Proxima b') with a minimum mass of $1.3\me$ orbiting within the classical habitable zone of our closest stellar neighbour, the M-dwarf Proxima Centauri \citep{Anglada2016}, raises important questions about how this planet formed and evolved into the object that we see today. The analysis presented in \citet{Anglada2016} gives an orbital period of 11.2 days and a formal upper limit on the eccentricity of $e \le 0.35$. There is also an as-yet unconfirmed hint in the data of an additional body with an  orbital period in the range 50--500 days. As they stand, these data provide little in the way of constraints on the formation history of Proxima b. Future observations that are able to place tighter constraints on the multiplicity of the Proxima planetary system, and on the orbital eccentricity and composition of Proxima b, would be extremely useful in constraining the formation mechanism.

Given the limited data currently available on the planetary system orbiting Proxima (from now on we refer to the host star as Proxima and the planet as Proxima b), our approach in this paper is to explore the consequences for planetary multiplicity, orbital eccentricity and water delivery that arise from a number of different formation scenarios for Proxima b. We discuss these scenarios in more detail below, but very briefly these can be described as follows: (i) \emph{In situ formation via giant impacts}. Here we use N-body simulations to examine the formation of Proxima b through mutual collisions between planetary embryos that orbit close to the current orbital location of Proxima b. The simulations are conducted under gas free conditions, as we assume this final stage of accumulation arises once the gaseous protoplanetary disc has dispersed; (ii) \emph{Formation via giant impacts between migrating embryos}. Here we use N-body simulations to model the growth of Proxima b analogues through mutual collisions between embryos that form across a wide range of orbital radii during and after the gas disc life time. Migration of planetary embryos due to gravitational interaction with the gas disc are included; (iii) \emph{Migration and accretion onto a single embryo}. Here we explore the formation of Proxima b analogues at large distance from Proxima via the formation of a single planetary embryo that forms in a protoplanetary disc and migrates while accreting planetesimals; (iv) \emph{Pebble accretion}. We have implemented the pebble accretion model of \citet{Lambrechts14} and explore the conditions under which a Proxima b analogue can form via this mechanism. Scenario (i) closely matches earlier work on the formation of the Solar System terrestrial planets \citep{ChambersWetherill1998}, but adopts significantly augmented disc masses to account for the fact that a reasonable model for the protoplanetary disc that would have existed around Proxima contains too little mass to form Proxima b directly. As such, this is similar to the in situ models of planet formation by \cite{HansenMurray2012} that have been proposed to explain the compact multiplanet systems discovered by Kepler. Scenario (ii) is closely related to earlier work by \citet{TerquemPapaloizou2007, McNeilNelson2010, Cossou14, Hellary, ColemanNelson14, ColemanNelson16} which combine accretion and migration. The study of scenario (iii) adopts simulations similar to \citet{Tanaka}, and the calculations conducted to study scenario (iv) are similar to those that have recently been presented by \citet{Lambrechts14} given that they are based on an implementation of their pebble accretion model. While the four scenarios that we explore do not cover the full range of possible formation histories, they do in some sense bracket the range of plausible scenarios and hence provide a useful guide of how the formation pathway relates to some of the present day properties of the system. 

Our aim is not to conduct simulations that result in a perfect match to the known orbital and physical parameters of Proxima b, but instead to demonstrate that each of the models that we consider are feasibly able to form a planet that is similar to Proxima b. For this reason, we introduce the notion of a Proxima b analogue that is defined to be a planet with mass in the range $0.75 \me \le m_{\rm p} \le 2 \me$ and orbital period in the range $6 \le P \le 20$ days. 

As mentioned above, \citet{Anglada2016} also detect a second signal in the period range 50--500 days, but its nature is deemed unclear due to stellar activity and inadequate sampling.
It is possible that this signal is due to a super-Earth orbiting within that period range, but could also be due to stellar variability.
If the signal is due to a super-Earth, then the stability of the system needs to be addressed.
Additional planets can cause Proxima b to have a non-zero eccentricity, influencing not only its orbital evolution, but also its internal properties by allowing continued tidal evolution between Proxima b and Proxima Centauri.
In this paper, as well as investigating possible formation scenarios for Proxima b, we also examine the dynamical stability of Proxima b's orbit in the presence of an additional super-Earth with an orbital period in the period range above.
We find that in a \emph{stress test} model, where the outer planet is placed in a 5:1 resonance with Proxima b, the planetary system survives for at least $10^7$ years with, however a significant interaction.
Since the \emph{stress test} is analogous of a `worst case scenario', we expect that if the additional planets had much longer periods, then the interactions will be significantly reduced.

This paper is organised as follows. We briefly describe the physical model in Section~\ref{sec:physicalmodel}, and in Section~\ref{sec:ProximaDisc} we describe the reference protoplanetary disc model that forms the basis of this study. We present the simulations for each of the four planet formation scenarios in Section~\ref{sec:simulations}. We include an additional planet to perform a dynamical study in Section~\ref{sec:dynamical}, and discuss our results in Section~\ref{sec:conc}.

\section{The physical model}
\label{sec:physicalmodel}

\begin{figure*}
\centering
\includegraphics[scale=0.28]{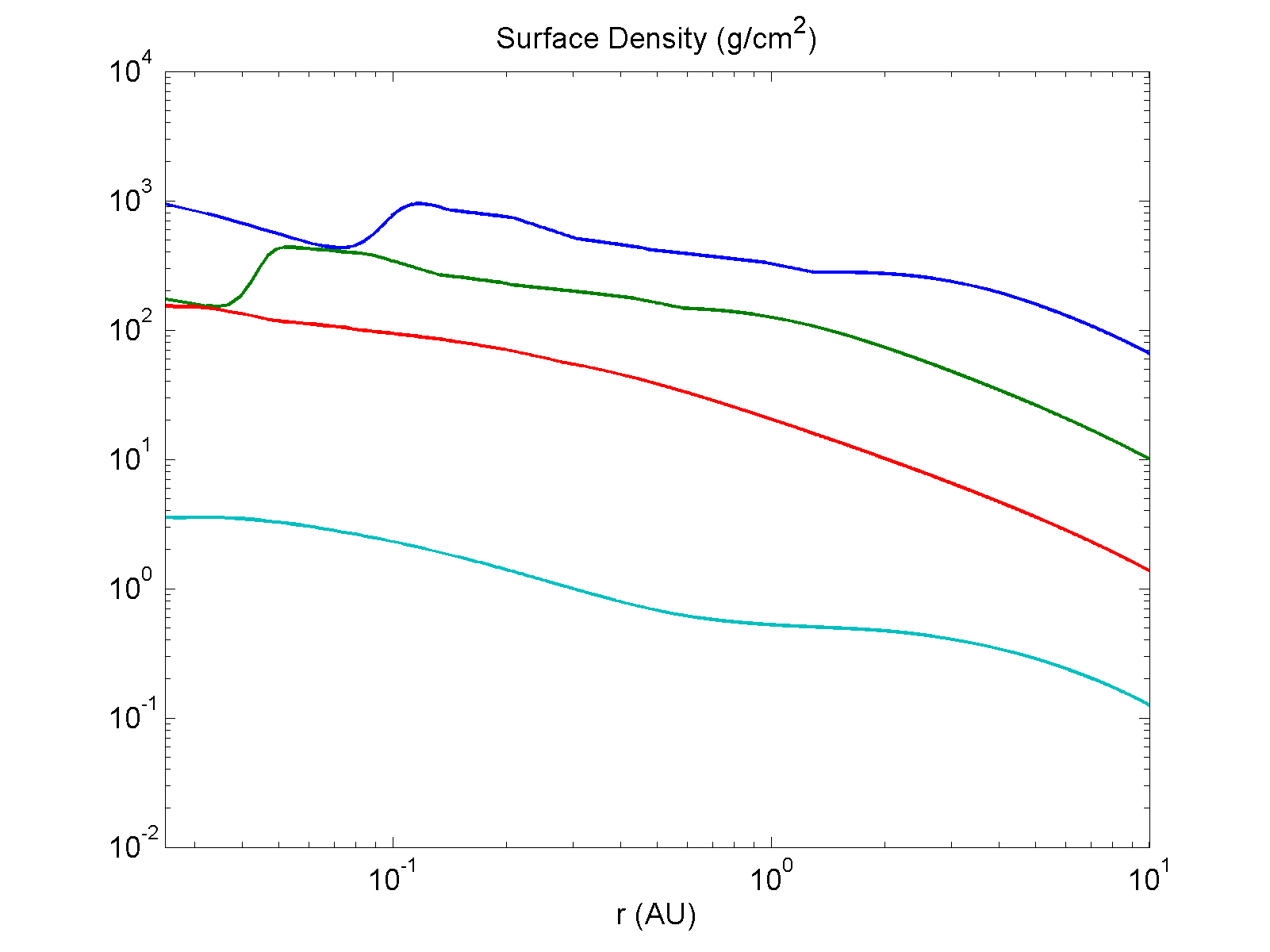}
\includegraphics[scale=0.28]{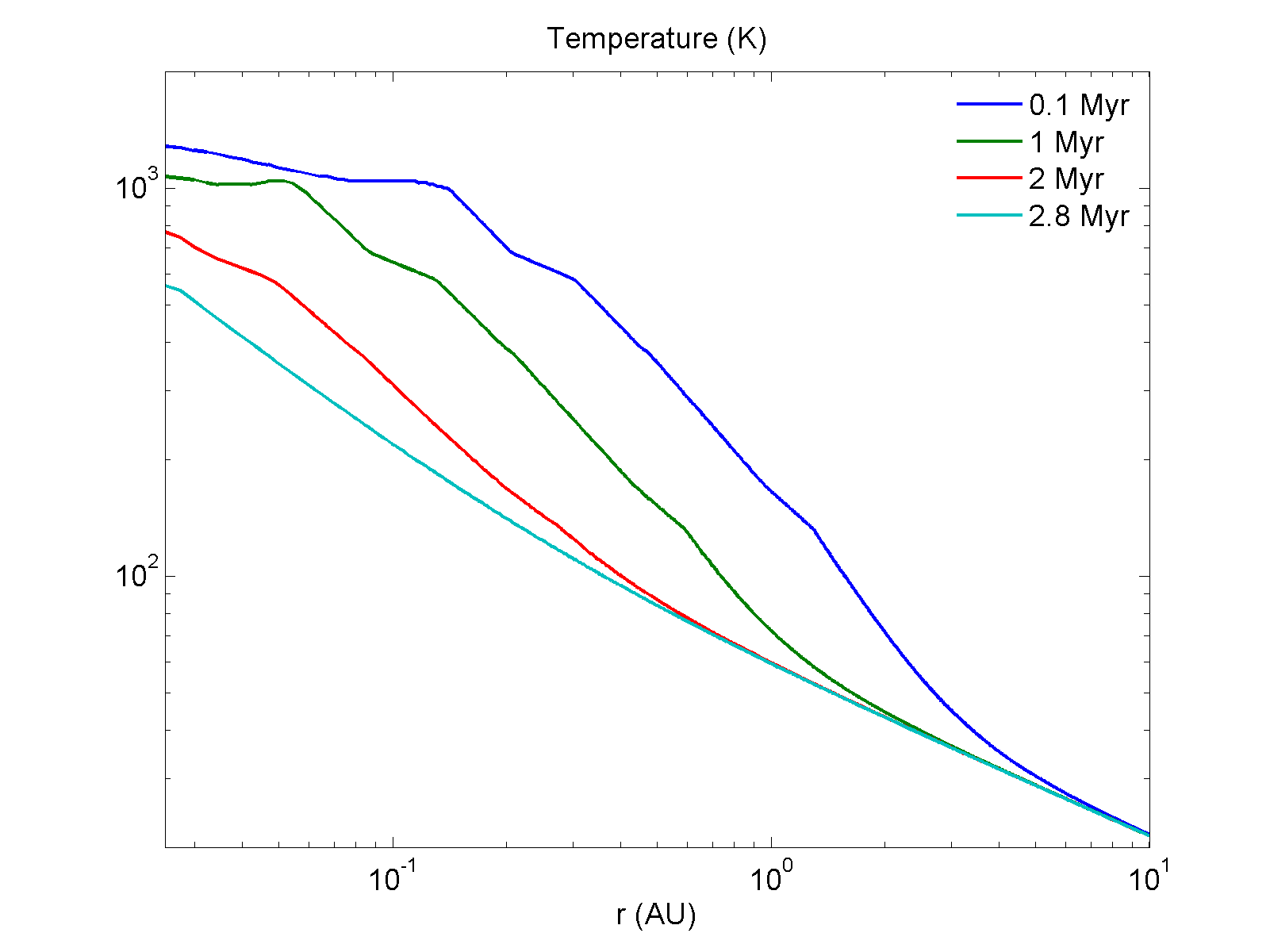}
\includegraphics[scale=0.28]{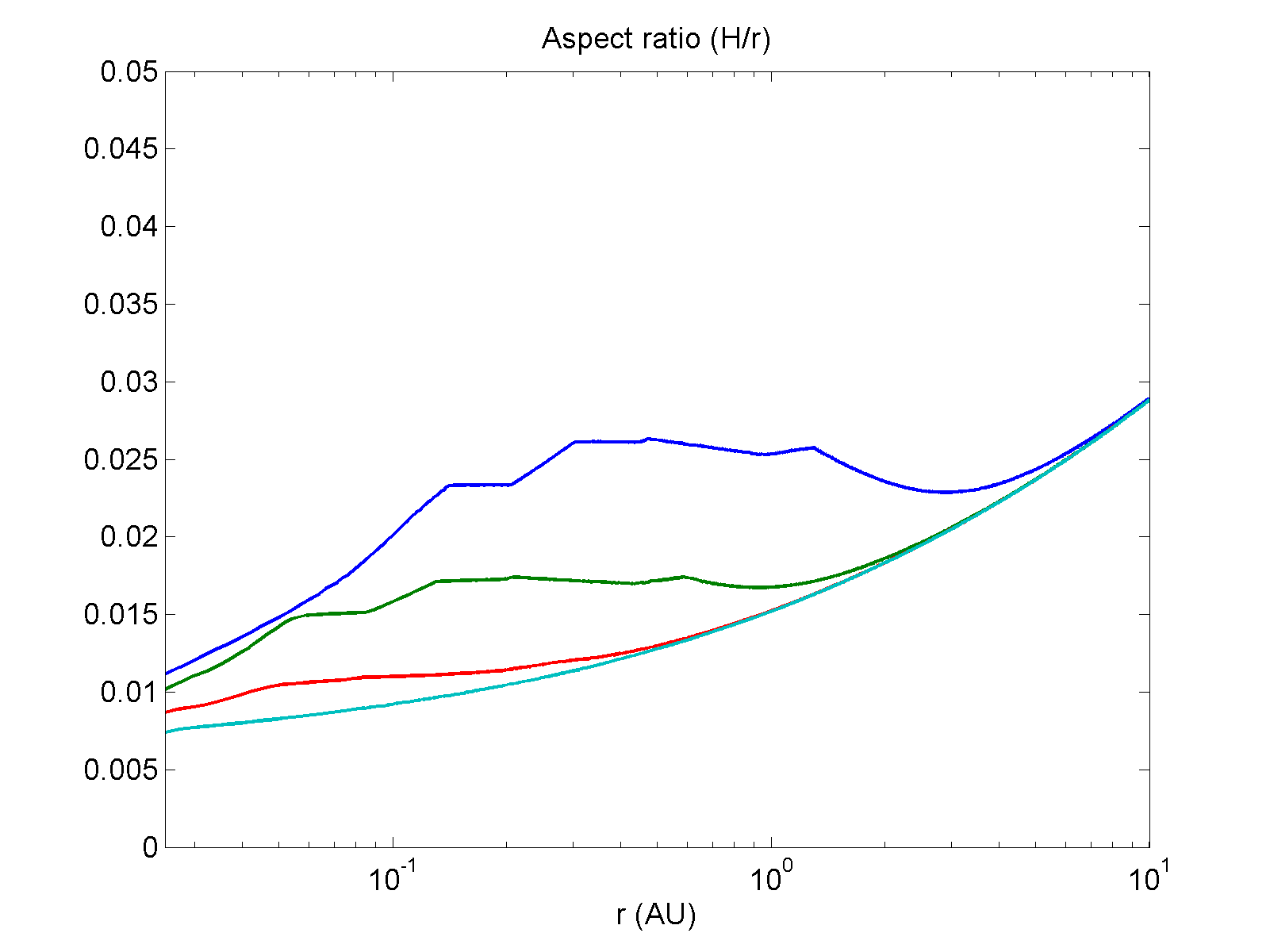}
\caption{Gas surface densities, temperatures and aspect ratios after 0.1, 1, 2 and 2.8 Myr (top-bottom lines)
of the standard Proxima protoplanetary disc (lifetime: 2.9 Myr).}
\label{fig:multiplot}
\end{figure*}

The physical model that we adopt for this study is based on the disc and planet formation models presented in \citet{ColemanNelson14, ColemanNelson16}. 
The N-body simulations presented here were performed using the Mercury-6 symplectic integrator \citep{Chambers}, 
adapted to include the disc models and physical processes described below. 

\noindent
(i) The standard diffusion equation for a 1D viscous $\alpha$-disc model is solved \citep{Shak,Lynden-BellPringle1974}. 
Temperatures are calculated by balancing black-body cooling against viscous heating and stellar irradiation. The viscous parameter $\alpha=2 \times 10^{-3}$ throughout most of the disc, but increases to $\alpha=0.01$ in regions where $T \ge 1000$ K to mimic the fact that fully developed turbulence can develop in regions where the temperature exceeds this value.
Although the above disc model has a moderate level of sophistication, we note that a number of recent studies present a different paradigm for disc evolution in which angular momentum transport is dominated by magnetised disc winds that are launched from the highly ionised disc surface regions throughout most of the radial extent to the disc \citep[e.g.][]{Bai2013,Gressel2015,Bai2016}. It is noteworthy that the global evolution of these discs, in terms of their surface density and thermal structure, is quite different from the ones that we consider in this work \citep[e.g.][]{Suzuki2016}, and would likely lead to planetary systems with quite different architectures and chemical abundance patterns than the ones that are produced in the simulations that we present here. There is clearly scope for future work that examines the implications of these disc models for the formation of Proxima b and other planetary systems.
 
\noindent
(ii) The final stages of disc removal occur through a photoevaporative wind.
A standard photoevaporation model is used for most of the disc evolution \citep{Dullemond}, corresponding to a photoevaporative wind being launched from the upper and lower disc surfaces. Direct photoevaporation of the disc is switched on during the final evolution phases when an inner cavity forms in the disc, corresponding to the outer edge of the disc cavity being exposed to the stellar radiation \citep{Alexander09}.
 
\noindent
(iii) The N-body simulations consist of one or more planetary embryos that can mutually interact gravitationally and collide.  In addition, some models also include small bodies that we refer to as boulders (bodies with radii $R_{\rm p} \sim 10$ m) or planetesimals (bodies with radii $R_{\rm p} \ge 100$ m). Planetesimals and boulders orbiting in the gaseous protoplanetary disc experience size dependent aerodynamic drag \citep{Weidenschilling_77}.
Collisions between protoplanets and other protoplanets or planetesimals resulted in perfect sticking. Planetesimal-planetesimal interactions and collisions are neglected for reasons of computational speed.
 
\noindent
(iv) We use the torque formulae from \citet{pdk10,pdk11} to simulate type I migration
due to Lindblad and corotation torques acting on the planetary embryos. Corotation torques arise from both entropy
and vortensity gradients in the disc, and the possible saturation of these torques is 
included in the simulations. The influence of eccentricity and inclination on the
migration torques, and on eccentricity and inclination damping are included
\citep{Fendyke,cressnels}.

\noindent
(v) The effective capture radius of planetary embryos accreting planetesimals is enhanced
by atmospheric drag as described in \citet{Inaba}.

The inner boundary of the computational domain is located
just inside of $0.025 \au$. Any planets whose semi-major axes are smaller than this boundary radius are removed from the simulation and are assumed to have hit the star. 

\section{A model for Proxima Centauri's protoplanetary disc}
\label{sec:ProximaDisc}

We have adopted a simple approach to constructing a fiducial model for the protoplanetary disc that could have surrounded Proxima during the epoch when Proxima b was forming. The minimum mass solar nebula model is often adopted when considering the formation of the Solar System's planets, and this contains $\sim 0.015\msun$ in a disc that extends out to 40 AU \citep{Hayashi}. Studies of the Solar System often augment this by a factor $\sim 3$ to account for the formation of the outer planets within time scales that agree with observed disc lifetimes \citep{Pollack}. Adopting a similar approach, we construct a \emph{Proxima disc} that contains 4.5\% of Proxima's mass. The radius of the disc depends on the initial angular momentum of the molecular cloud that collapsed to form the star, and this is obviously an unknown quantity. \cite{WilliamsCieza11} quote a relationship between disc mass and the radius, $R_{\rm c}$, at which the surface density changes from a power-law distribution to an outer exponential tail, based on model fits to observations: $M_{\rm d} \propto R_{\rm c}^{1.6 \pm 0.3}$. Assuming a linear scaling between stellar mass and disc mass, we invert this relation to obtain an approximate radius of Proxima's disc based on Proxima's mass, the Solar mass and the radius of the Solar System (i.e. 40 AU). This gives a radius for Proxima's disc of $\sim 10$ AU.
One source of uncertainty associated with this value is the adopted radius of the Solar System. The Solar System's protoplanetary disc could have been much larger than the current size of the Solar System, meaning that our radius of $10 \au$ for Proxima's disc should be interpreted as a minimum estimate.

Stellar evolution models that compute Hayashi tracks for contracting low mass protostars indicate that the luminosity of Proxima at an age of 1 Myr would be $\sim 0.1$ L$_{\odot}$ \citep{Stahler2005}. Combined with the viscous dissipation that arises in the gas discs that we adopt, the snowline where $T < 170$ K occurs at $\sim 1 \au$ at the start of the simulations. The metallicity of Proxima is measured to be ${\rm [Fe/H]}=0.21$ \citep{Anglada2016}. When considering the distribution of solids, our baseline assumption is that interior to the snowline the dust-to-gas ratio is $0.005 \times 10^{0.21}$, where the Solar value $\sim 0.005$. Following \citet{Lodders2003}, we assume that the dust-to-gas ratio increases by a factor of $\sim 2$ beyond the snowline due to the condensation of water.
A disc with a power-law surface density profile given by $\Sigma(R) \propto R^{-1/2}$ then has $1.5 \me$ of solids distributed interior to 1.6 $\au$ and a total solids mass of $\sim 25 \me$ across the entire disc, indicating that substantial mobility of solids is required to form a planet with Proxima b's minimum mass orbiting at a distance of $\sim 0.05 \au$ from the star.

\subsection{Evolution of the Proxima protoplanetary disc}

Figure \ref{fig:multiplot} shows the evolution of the Proxima disc, which has a lifetime of 2.9 Myr.
Disc surface density profiles are shown in the left panel, temperature profiles in the middle panel, and $H/r$ profiles in the right panel.
The times corresponding to each profile are indicated in the middle panel.
As time progresses the gas disc viscously accretes onto the central star, gradually reducing the surface density and temperature over time as illustrated by the different profiles in fig \ref{fig:multiplot}.
The dips in the surface density profiles close to the inner edge of the disc ($r\le 0.1\au$) are a result of the active turbulent region, where T$>1000$K causes an increase in the viscosity.
Over time, this region moves in towards the star as the reduction in surface density reduces the viscous heating rate and opacity.
The turbulent region disappears when the disc temperature no longer exceeds 1000 K anywhere in the disc, as is shown by the red line in fig. \ref{fig:multiplot}. 
The dip in the surface density in the bottom cyan line shows the onset of disc dispersal, where the photoevaoporation rate is larger than the viscous rate.
After this occurs, the disc quickly disperses from the inside out within $\sim 0.1$ Myr.

\begin{figure}
\centering
\includegraphics[scale=0.45]{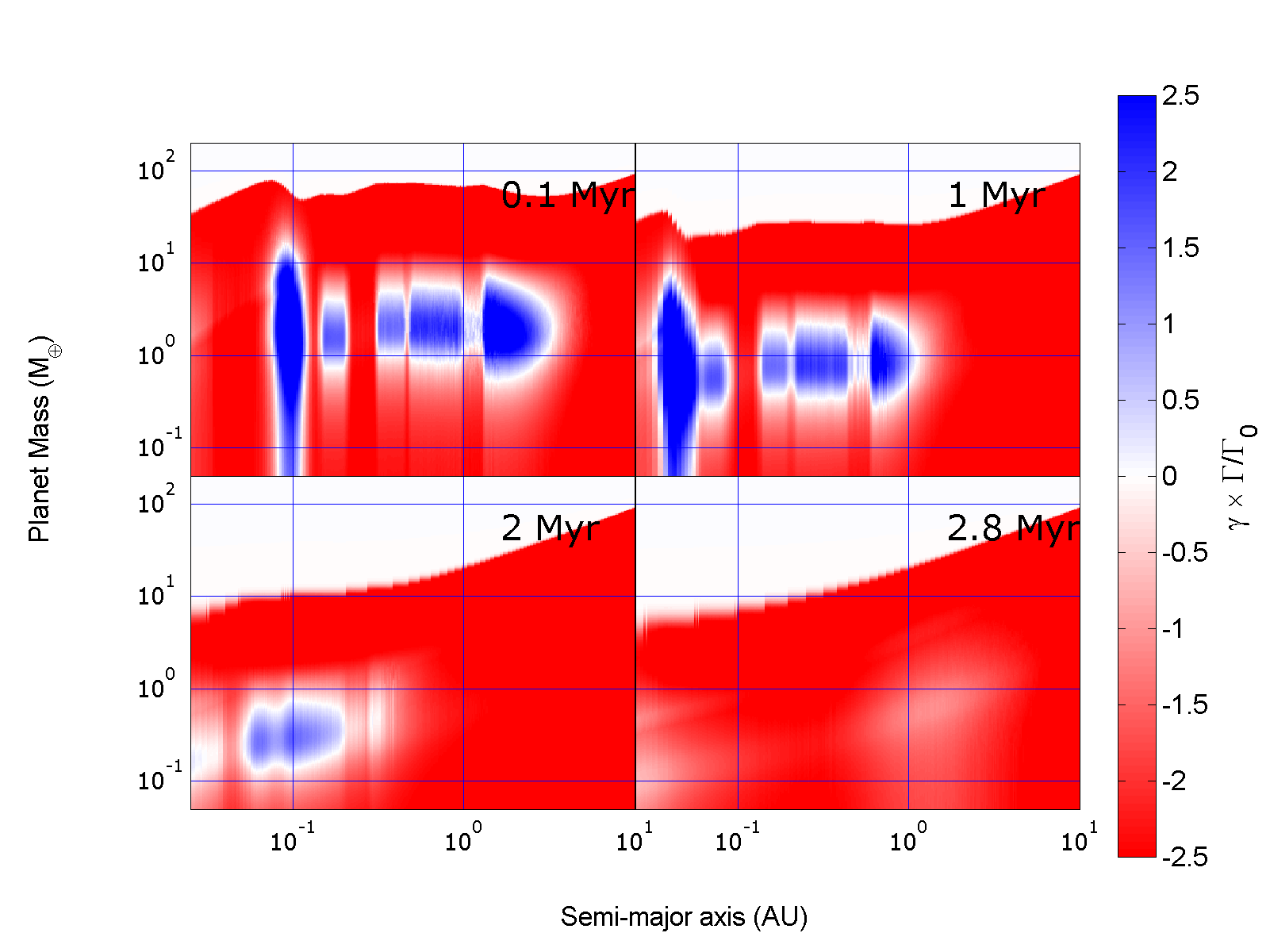}
\caption{Contour plots showing regions of inwards (red) and outwards (blue) migration) in the Proxima disc at t = 0.1 Myr (top left), 1 Myr (top right), 2 Myr (bottom left) and  2.8 Myr (bottom right).}
\label{fig:contours}
\end{figure}

Type I migration of planets is controlled by both Lindblad and corotation torques.
In our disc models Lindblad torques are negative and corotation torques are generally positive.
Strong, positive corotation torques arise in regions where the radial entropy gradient is negative, and this is usually the case in the inner disc regions where viscous heating dominates over stellar irradiation.
Corotation torques may saturate when either the viscous or thermal time scale differs significantly from the periods of horseshoe orbits executed by gas in the corotation region.
Figure \ref{fig:contours} shows migration contours that illustrate the migration behaviours of planets as a function of their masses and semi-major axes in a standard Proxima disc.
Dark blue regions correspond to strong outward migration, red regions correspond to strong inward migration, and white contours represent regions where Lindblad torques balance corotation torques, often referred to as \emph{zero-migration zones}.
The active turbulent region creates a planet trap in the interior regions of the disc in the first two panels of fig. \ref{fig:contours}.
This trap is formed due to the large positive gradient in surface density in those regions as is shown by the blue and green profiles in fig. \ref{fig:multiplot}.
Once the active turbulent region disappears, there are no longer any planet traps for planets of mass $m_{\rm p} > 1 \me$, as is shown by the bottom panels in fig. \ref{fig:contours}.
The outward migration regions that are present there also only extend out to $\sim 0.3 \au$.
These values have important implications on the evolution of planets in those systems.
Planets that are more massive than $1 \me$ are able to migrate inwards without reaching a zero-migration zone, whilst planets less massive than $1 \me$ will migrate in to zero-migration zone located at 0.3 $\au$, before slowly migrating in with the zone as the disc evolves and cools, which reduces the strength of the corotation torque.

The location of the snowline is important in determining not only the distribution of solids, but also their composition.
Solid material located exterior to the snowline is rich in volatiles such as water, whilst solids located interior to the snowline are rocky, devoid of volatiles.
This difference in locations and compositions is important when discussing the final composition of any Proxima b analogue.
At the start of the disc lifetime, the snowline is located at $\sim 1 \au$, such that any solid material exterior to $1 \au$ is assumed be volatile rich whilst any material interior to $1 \au$ will be rocky.
As the disc evolves, and gas surface densities and temperatures decrease, the snowline drifts inwards \citep{Oka2011,Mulders2015}.
At the end of the disc lifetime, it is located at $\sim 0.2 \au$, close to its expected location today ($0.1 \au$). 

\section{Formation Scenarios}
\label{sec:simulations}
We examine four different scenarios for the formation of Proxima b.
These include in situ formation of one or more planets from a collection of planetary embryos and planetesimals, formation and migration of multiple planetary embryos into the inner disc before colliding and forming  a Proxima b analogue, formation of a planetary embryo before migrating through a swarm of planetesimals resulting in a planet similar to Proxima b, and formation of a planet through pebble accretion while migrating close to the central star.
To investigate all scenarios, we used the Mercury-6 symplectic integrator \citep{Chambers}, and where a protoplanetary disc is required, we utilised the thermally-evolving viscous disc model of \citet{ColemanNelson16}, which includes up-to-date prescriptions for disc evolution, photoevaporation, gas drag and planet migration as described in Section~\ref{sec:physicalmodel}.

\begin{figure*}
\includegraphics[scale=0.42]{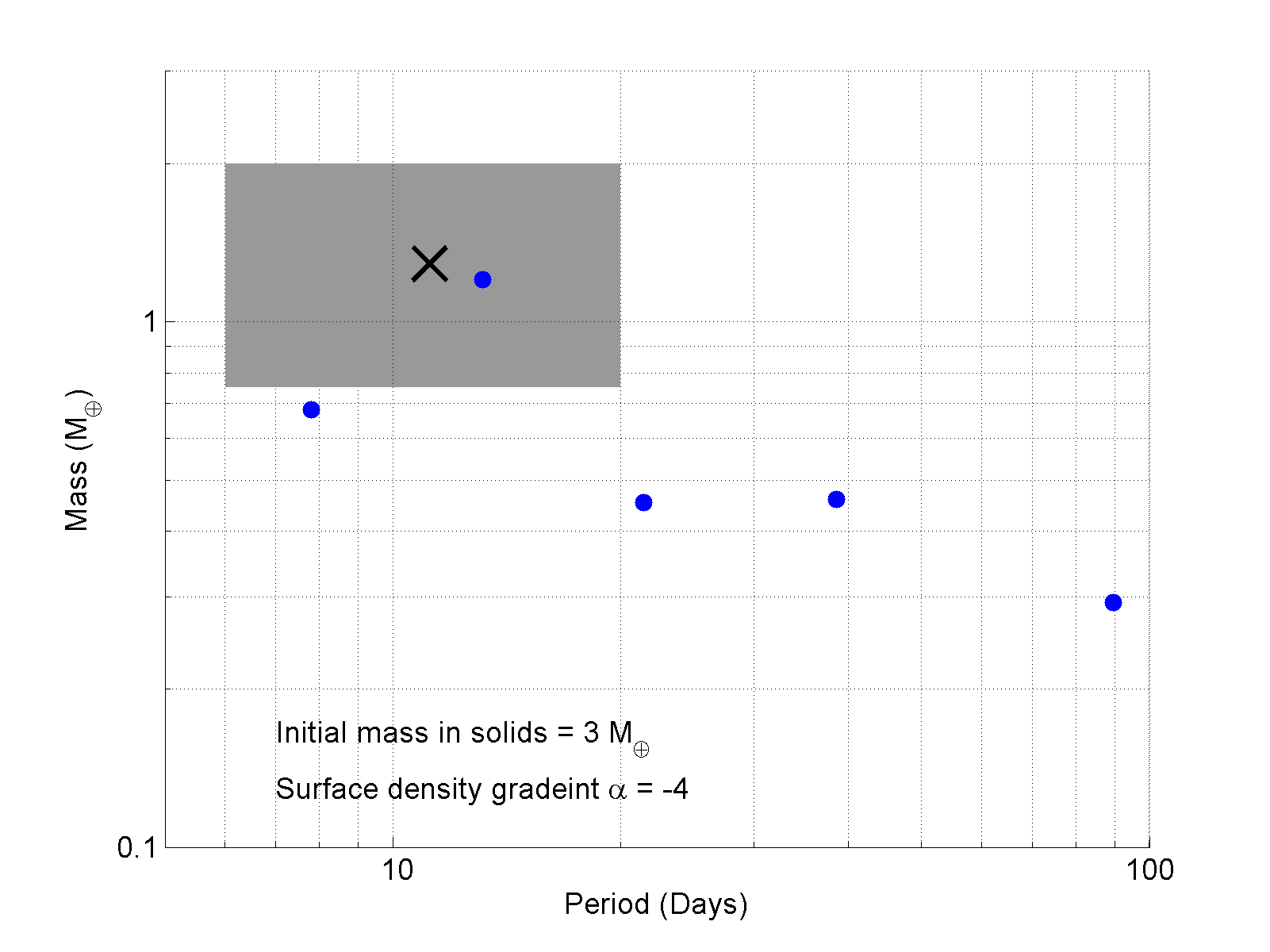}
\includegraphics[scale=0.42]{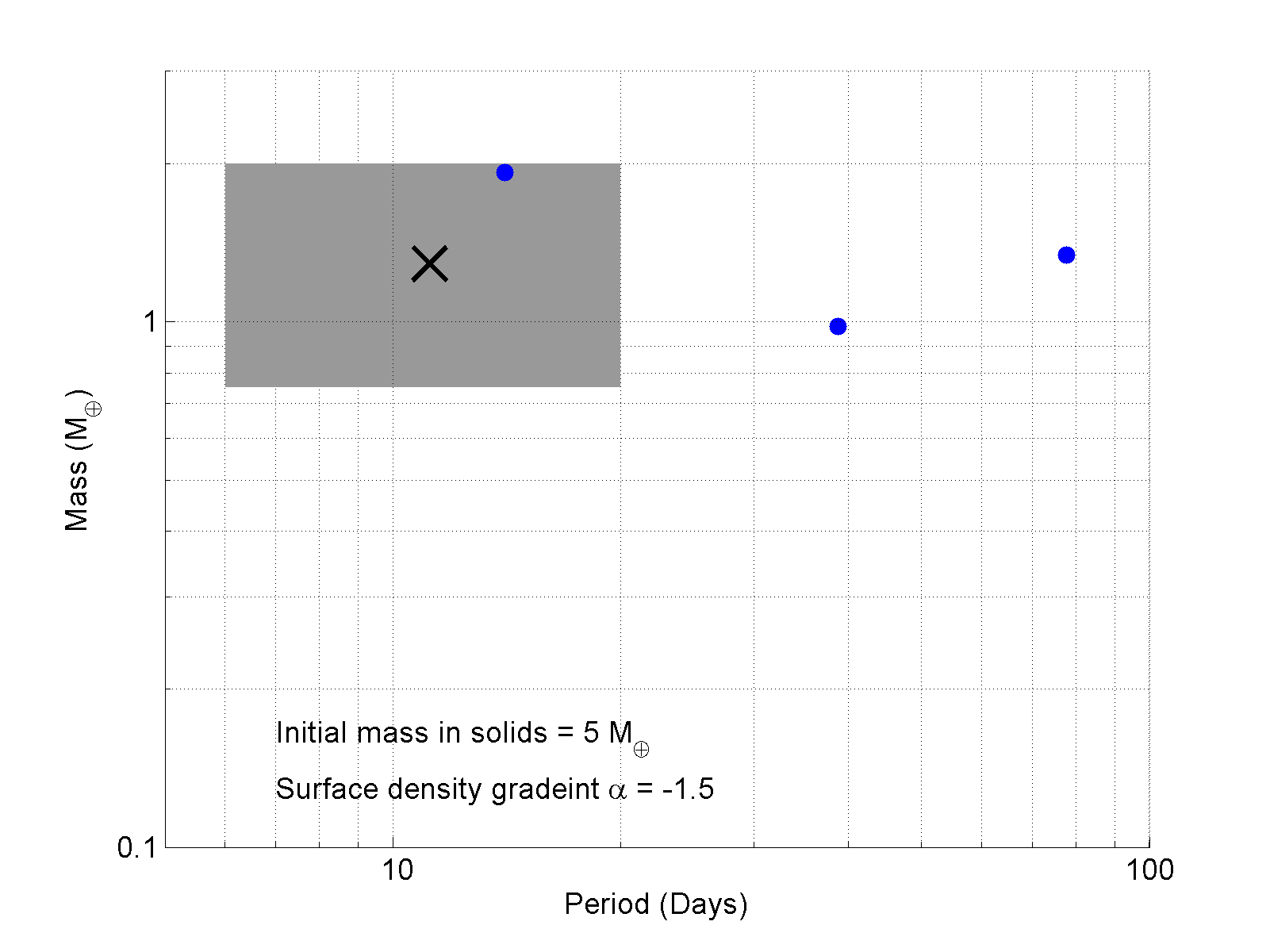}
\includegraphics[scale=0.42]{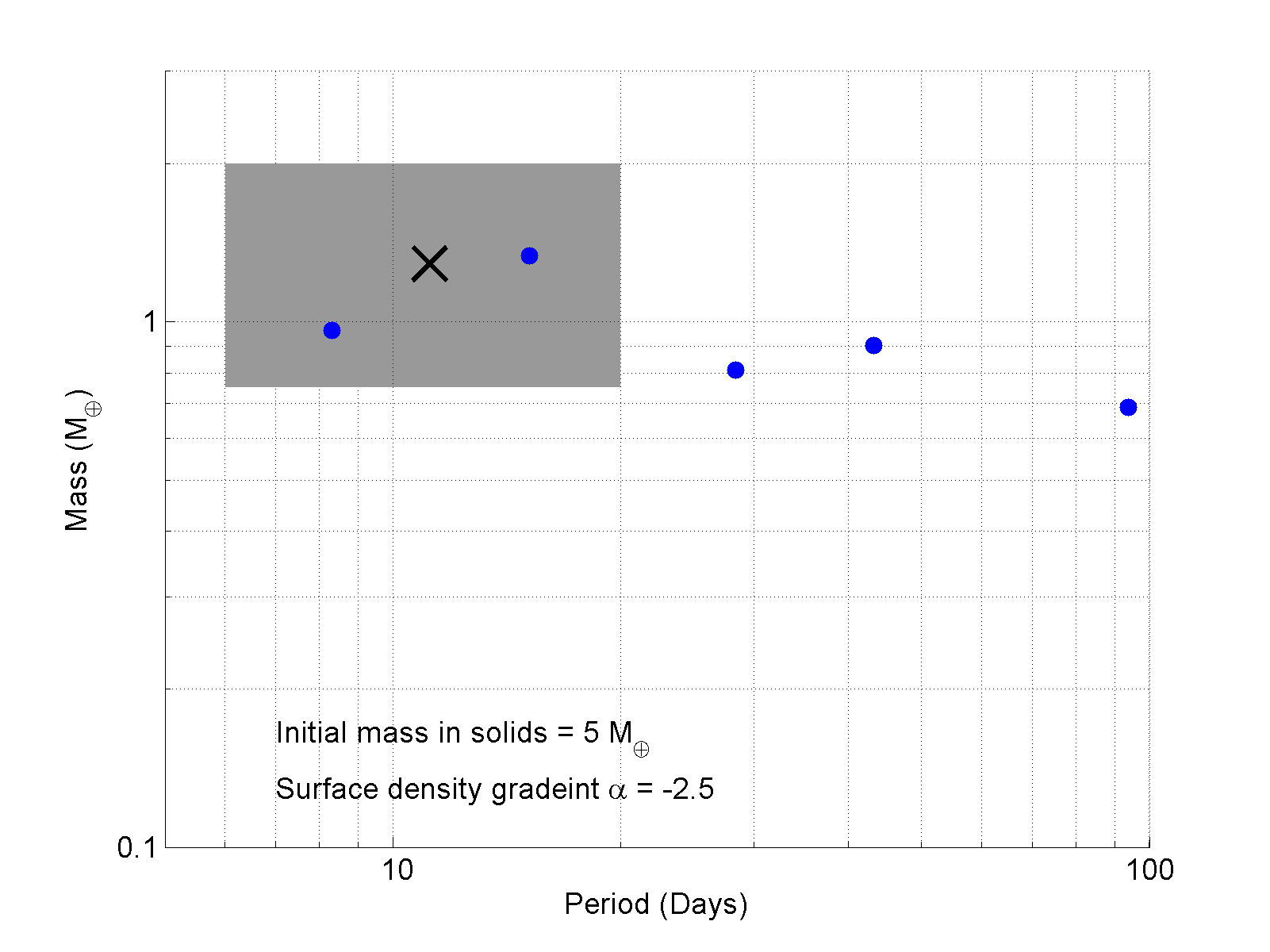}
\includegraphics[scale=0.42]{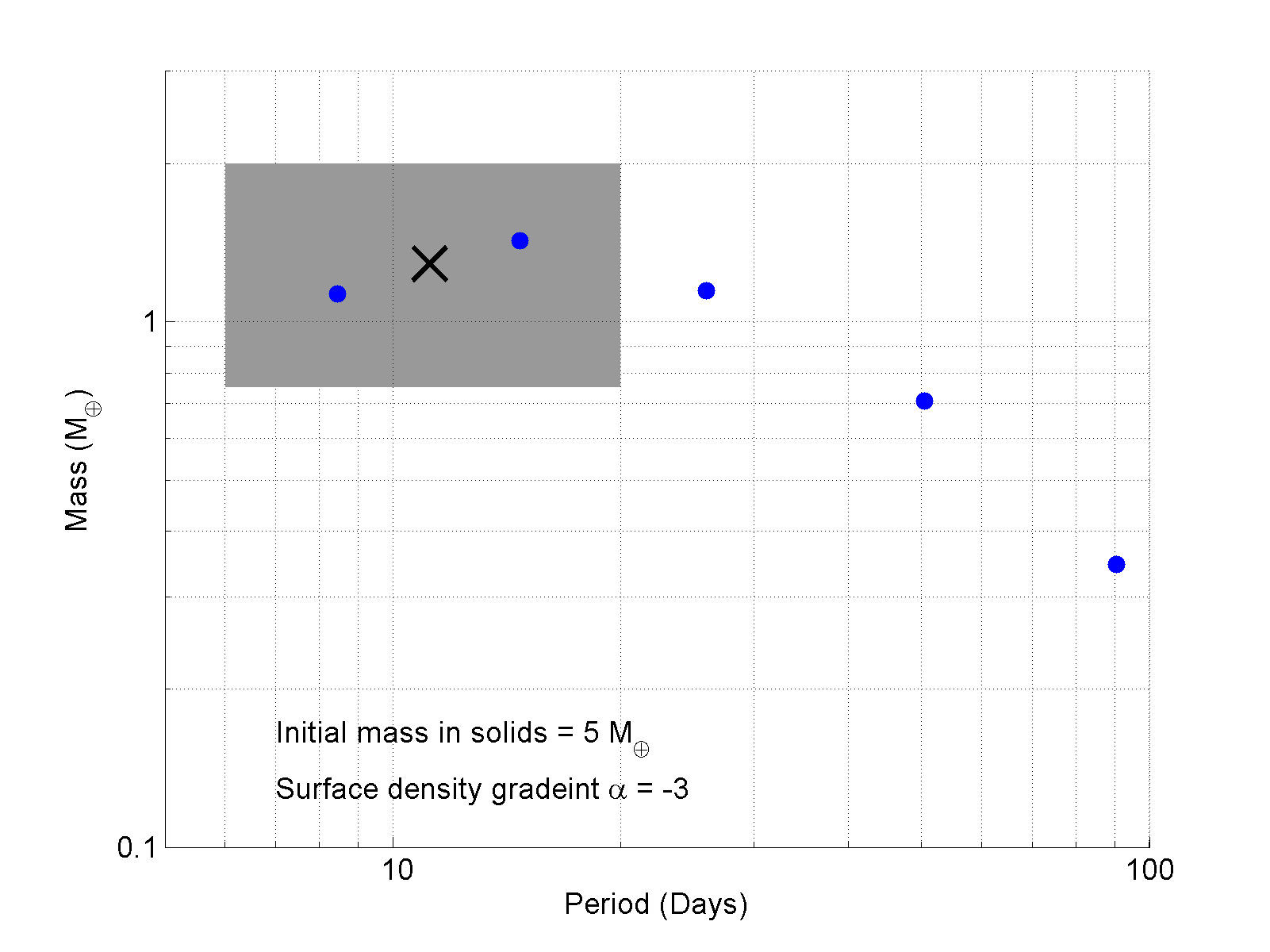}
\caption{A mass versus period plot showing the final masses and periods for simulated planetary systems in the \emph{in situ} formation scenario (blue dots). The shaded grey region represents the mass and period range that planets can be classified as Proxima b analogues, and the black cross represents Proxima b.}
\label{fig:terrestrial}
\end{figure*}

\subsection{In situ planet formation}
\label{sec:terrestrial}
The terrestrial planets in the Solar System are thought to have formed in situ, that is they formed at/or close to their current orbital positions \citep{ChambersWetherill1998}.
To form in situ, they required numerous less massive embryos embedded within a swarm of smaller planetesimals to interact and collide in a gas free environment, resulting in the majority of the original mass in the disc of solids being incorporated within the terrestrial planets seen today.
We adapt this scenario here, by running numerical simulations to study whether a Proxima b analogue can form close to Proxima through the mutual accretion of planetary embryos embedded in a disc of planetesimals.

We begin by placing a number of planetary embryos with masses, $m_{\rm p} = 0.05 \me$ within a disc of planetesimals, whose individual masses are 20 times less that of the protoplanets.
We set the total mass in solids to be 3, 5 or 10 $\me$, and we distribute the planets and planetesimals between 0.04 and 0.25 $\au$ using the following surface density profile:
\begin{equation}
\Sigma_{\rm s}(r) = \Sigma_0(1\au)\left(\frac{r}{1\au}\right)^{\alpha}
\end{equation}
where $\alpha$ is assigned a value between -0.5 and -4, taken at intervals of 0.5.
Since there is insufficient solid material initially located in the inner regions of the protoplanetary disc that likely existed around Proxima, an implicit model assumption is that the solid material accumulated to form small bodies further out in the disc, before migrating into the inner disc regions via aerodynamic drag over the disc lifetime, where it then accumulated to form the planetesimals and protoplanets that provide our initial conditions. In recent work, \citet{Ogihara15} show that low mass embryos located in the inner disc are likely to grow rapidly through mutual collisions while in the presence of the gas disc, and then type-I migrate in to the inner edge of the disc, calling into question the realism of our initial conditions. One possibility is that MHD turbulence in the hot inner disc regions could slow the growth rates of the embryos such that the rapid collisional growth observed by \citet{Ogihara15} is slowed appreciably. We note that we assume that the embryos and planetesimals will be volatile-free, since icy material in the form of small bodies that migrates interior to the snowline would sublimate leaving only refractory material to continue migrating and accumulate at short orbital periods. A potential problem with this assumption is that the ice-line migrates inwards as the disc evolves and cools, as discussed above and in numerous other works \citep[e.g.][]{Oka2011,Mulders2015}. In principle this can lead to initially dry material in the inner regions of the disc becoming mixed with inward drifting icy particles that move into the inner disc with the moving ice-line, such that the accreting material is volatile rich. This issue is important for the terrestrial planets of the Solar System, which are highly depleted in volatiles even though the ice-line would have moved into $\sim 1 \au$ as the protosolar nebula evolved. \cite{Morbidelli2016} suggest that the dryness of the Earth, Mars and inner asteroid belt may be because the inward drift of icy particles was prevented by the presence of proto-Jupiter. In our model the ice-line does not migrate in as far as the current location of Proxima b, so we would not expect it to be composed primarily of icy material, but it does reach the outer edge of the belt of bodies between $0.2$ and $0.25 \au$, and so these could in principle become rich in volatiles. Dynamical diffusion of these into the vicinity of Proxima b's orbital location could then lead to the delivery of moderate levels of volatiles.

In principle, the prompt growth of protoplanets with masses $\sim 0.05 \me$ beyond the ice-line could allow delivery of volatile-rich bodies into the inner regions by type I migration before dispersal of the gas disc, as these bodies would not sublimate as they migrate through the ice-line. The type I migration time scale of $0.05 \me$ bodies located at $1 \au$ is $\sim 1$ Myr in our standard Proxima disc, assuming that it does not evolve with time. We have examined how far a $0.05 \me$ body can migrate from an initial location of $1 \au$ in our evolving disc model and find that it reaches a semi-major axis of $0.45 \au$ as the disc disperses, indicating that the ring of material that provides our initial conditions can not contain icy protoplanets that have migrated in from beyond the ice-line.

At the beginning of the simulations, planetary embryos start to accrete planetesimals in their local vicinity, whilst simultaneously dynamically stirring others.
This dynamical stirring increases planetesimal and planetary embryo eccentricities, increasing the frequency of close encounters and mutual collisions.
After 1 Myr, the majority of planetesimals have either been accreted by the planetary embryos or ejected from the system.
During this time, mutual collisions between planetary embryos also occurred, creating more massive planetary embryos that were able continue accreting material.
We stop the simulations after 100 Myr, at which point only a few planetary embryos remain in stable orbital configurations, having accreted/ejected all other planetary embryos and planetesimals in the simulations.

Figure \ref{fig:terrestrial} shows the final masses and periods for planetary embryos that formed and survived in four of the simulations.
The shaded grey areas represent the mass and period range required for a simulated planet to be classified as a Proxima b analogue, with the black cross representing the minimum mass and period of Proxima b.
We recall that we require a Proxima b analogue to have a mass between 0.75 and 2 $\me$ and an orbital period between 6 and 20 days.
It can be seen that a number of the surviving embryos match the characteristics (mass and orbital period) of Proxima b quite well.
These planets accreted the majority of their material from their local vicinities.
Since this material originated interior to the snowline, the composition of these planets will be rocky and dry.
For these planets to have any volatile material in their compositions, volatile-bearing planetesimals would have to be transported into their local vicinity.
As there is no gas disc to effect aerodynamic drag, this material has to undergo dynamical diffusion, i.e. it has to be scattered in through gravitational interactions with embryos from beyond the snowline.
If Proxima b analogues are able accrete material in this manner, then they would still be predominantly rocky, but would also contain modest amounts of volatiles. We have examined the feeding zones of the Proxima b analogues that form, and find that planetesimals initially located close to the outer edge of the disc of solids at $\sim 0.25 \au$ are able to accrete onto the planets. Although we do not consider discs of solids that extend beyond the snowline at $\sim 1 \au$ in this work, our simulations indicate that dynamical diffusion operates sufficiently efficiently in the models for volatile-rich material that could be located beyond the snowline in an extended disc to diffuse into the vicinity of the Proxima b analogue planets. Previous studies have also shown that dynamical diffusion of planetesimals operates strongly in discs which are sufficiently massive for Earth-mass bodies to form \citep{Raymond2007}, and similar studies also show that low mass discs (such as the Proxima disc introduced above) do not support large scale diffusion, such that systems of very low mass planets around M-dwarfs may be very poor in volatiles. The one caveat that needs to be considered here, however, is that the discs of solids considered here have been significantly augmented in mass through a supposed inward drift of material, so extending the solids disc out to beyond the snowline while maintaining a large surface density may not be a sensible thing to consider.

It is also noticeable that the surviving planets that formed in the simulations are not alone, in that there are numerous other planets orbiting close-by.
This multiplicity is a common outcome of the in situ formation scenario, where there are typically one or two planets containing a significant bulk of the surviving mass, accompanied by a number of less massive planets orbiting nearby as is shown in fig. \ref{fig:terrestrial}.
A significant number of the accompanying planets would be difficult to detect since they would only induce small radial velocities, $v_{\rm r}\leq 1 \rm {ms^{-1}}$, below current detection thresholds.
If Proxima b formed in situ from a collection of planetary embryos and planetesimals, then it can be expected that future observations could confirm the existence of other planets in the system.

\begin{figure}
\includegraphics[scale=0.45]{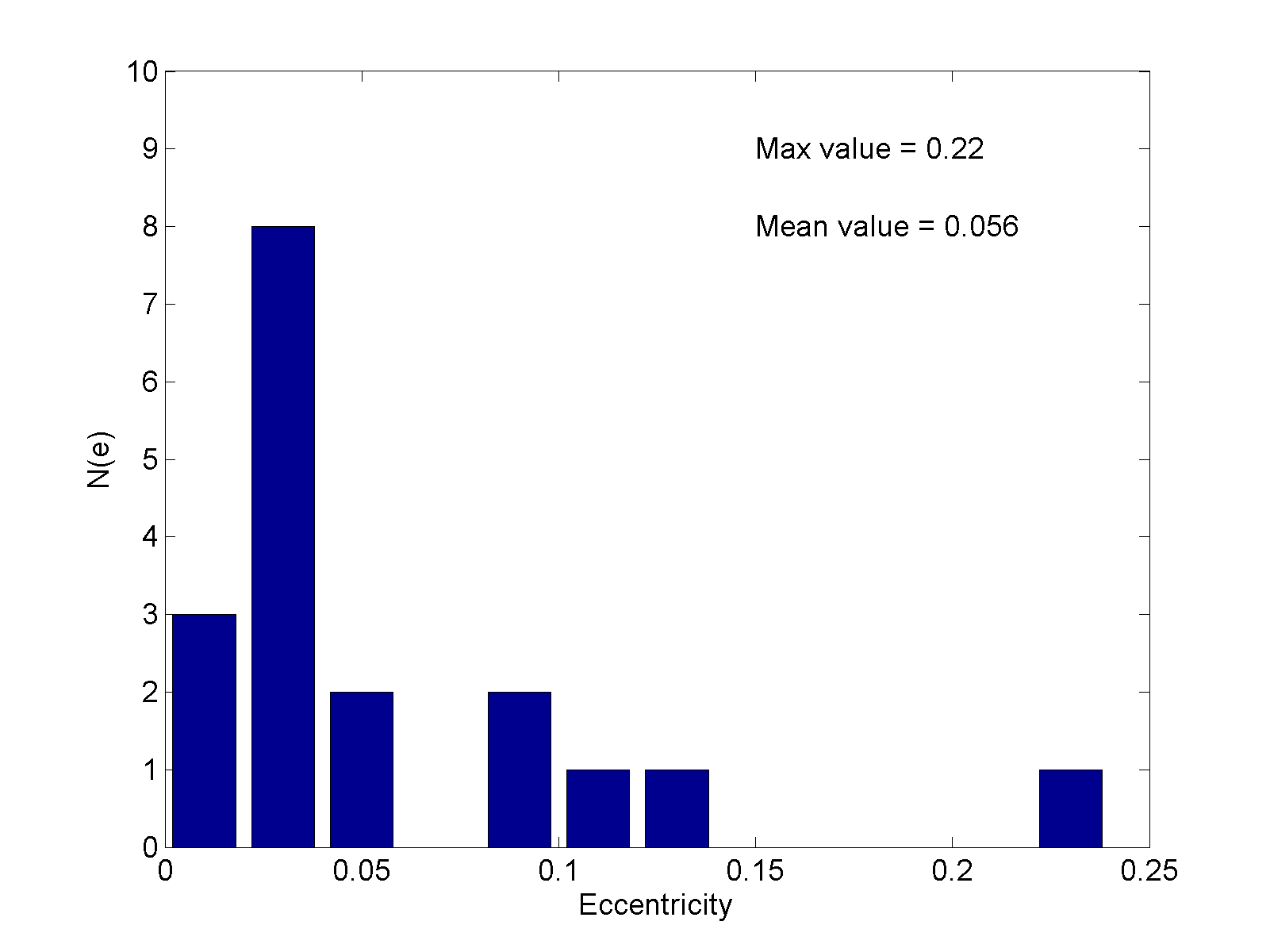}
\caption{Eccentricity distribution of Proxima b analogues formed in the in situ formation scenario. We define a Proxima b analogue as a planet with mass $0.75 \leq m_{\rm p} \leq 2\me$, and orbital period $6\leq P \leq20$ days.}
\label{fig:terrestrialecc}
\end{figure}

Another effect of multiplicity would be that the observed planets would possess non-zero eccentricities, even though tidal forces that act on the planets in proximity to their host stars would attempt to circularise their orbits. \citet{Ribas16} consider the tidal evolution of Proxima b, and show that for reasonable tidal dissipation rates, tidal circularisation is weak and operates on time-scales longer than the estimated 4.8 Gyr age of the star.
Figure \ref{fig:terrestrialecc} shows the eccentricity distribution of all simulated Proxima b analogues.
The average eccentricity is $\sim0.056$, and the maximum eccentricity obtained among all runs is equal to 0.22. As discussed below, this has implications for expectations concerning the tidal evolution of the planetary spin.
If the eccentricity of Proxima b could be measured accurately and it was shown to be significantly different from zero, this could imply the existence of other planets in the Proxima system.

\begin{figure}
\includegraphics[scale=0.45]{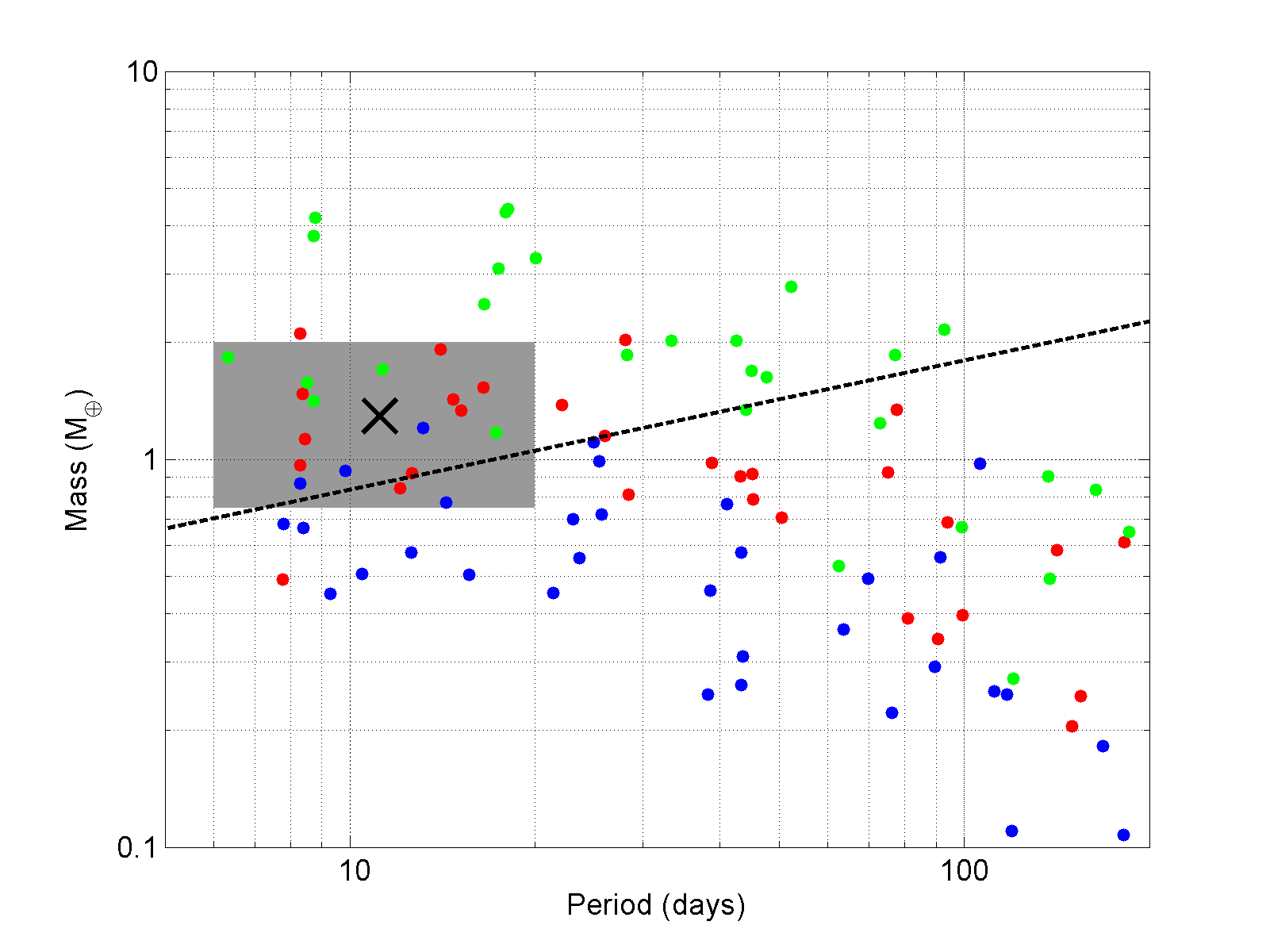}
\caption{A mass versus period plot for all surviving planets in the in situ formation scenario. Different colours represent the initial total mass in planetary embryos and planetesimals: blue - 3 $\me$, red - 5 $\me$, green - 10 $\me$. The shaded grey region represents the mass and period range of planets that are classified as Proxima b analogues, and the black cross represents Proxima b. The dashed line represents a radial velocity detection threshold of K$\sim1$ m/s.}
\label{fig:mvpterrestrial}
\end{figure}

Figure \ref{fig:mvpterrestrial} shows the final mass versus period distribution of all planets formed in this scenario.
It highlights that there are a significant number of Proxima b analogues, as well as a large number of less massive planets that would be difficult to detect, in addition to companion planets that would be above a detection threshold of 1 m/s, as shown by the planets with masses above the dashed line in fig. \ref{fig:mvpterrestrial}. It is interesting to note that simulations that contained a total solids mass of 3$\me$ occasionally only have a single planet that lies above the radial velocity detection limit, with the majority of companions being below the detection threshold. Simulations with a total solids mass of 10$\me$ however have numerous planets per simulation that are above the detection threshold, however these simulations also form very few Proxima b analogues as a result of the increased mass in the system. The simulations with total solids mass of 5$\me$ are able to form numerous Proxima b analogues as well as a number of other detectable and undetectable planets in the systems.

It is expected that in this scenario the Proxima b analogues would have negligible atmospheres in terms of mass.
Since they would have formed after the end of the gas disc lifetime, they would not harbour a primordial atmosphere abundant in hydrogen and helium.
Instead the planets could have acquired secondary atmospheres after their formation, through outgassing of volatiles from the planetary interior.

\begin{figure*}
\includegraphics[scale=0.42]{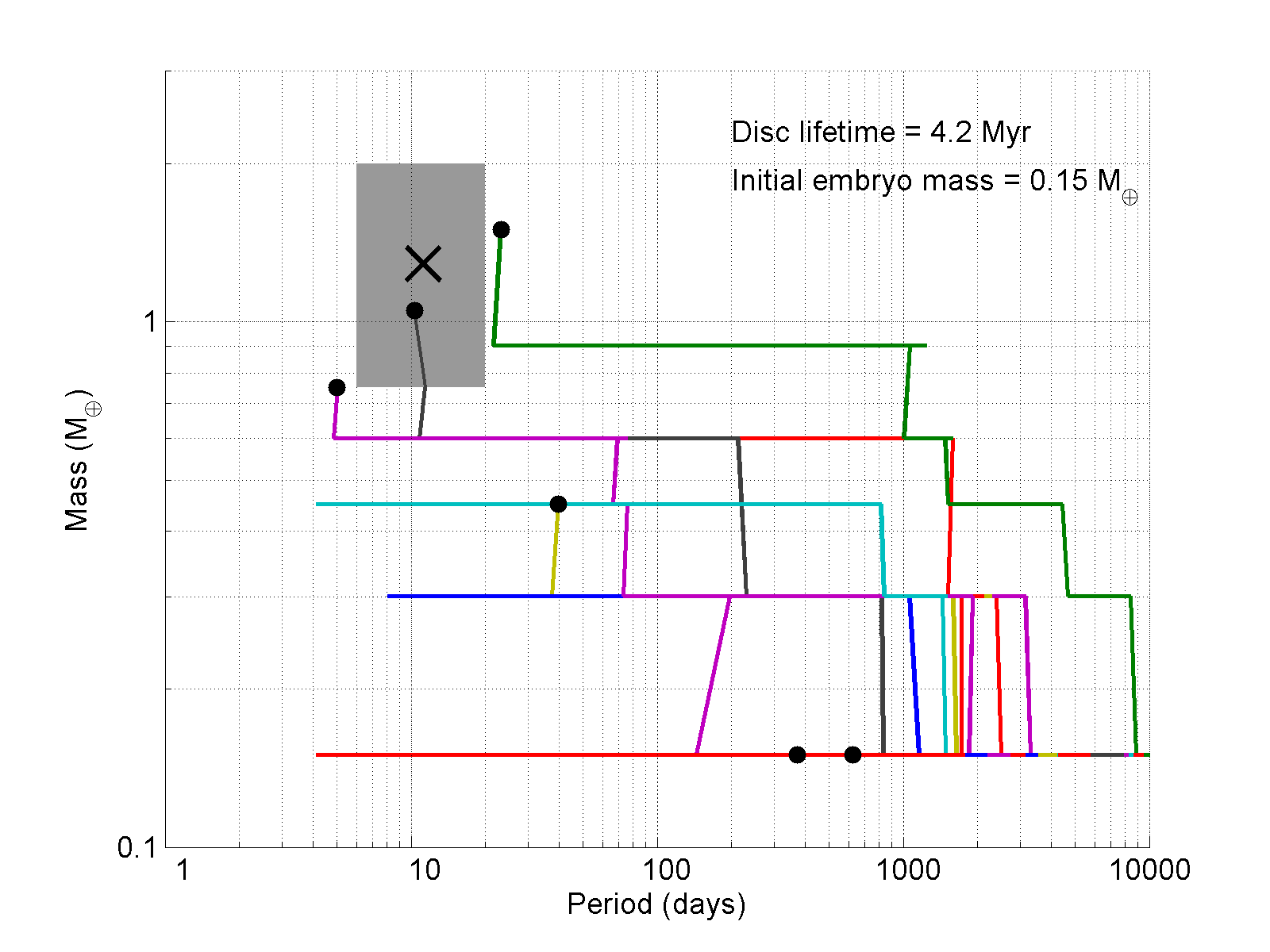}
\includegraphics[scale=0.42]{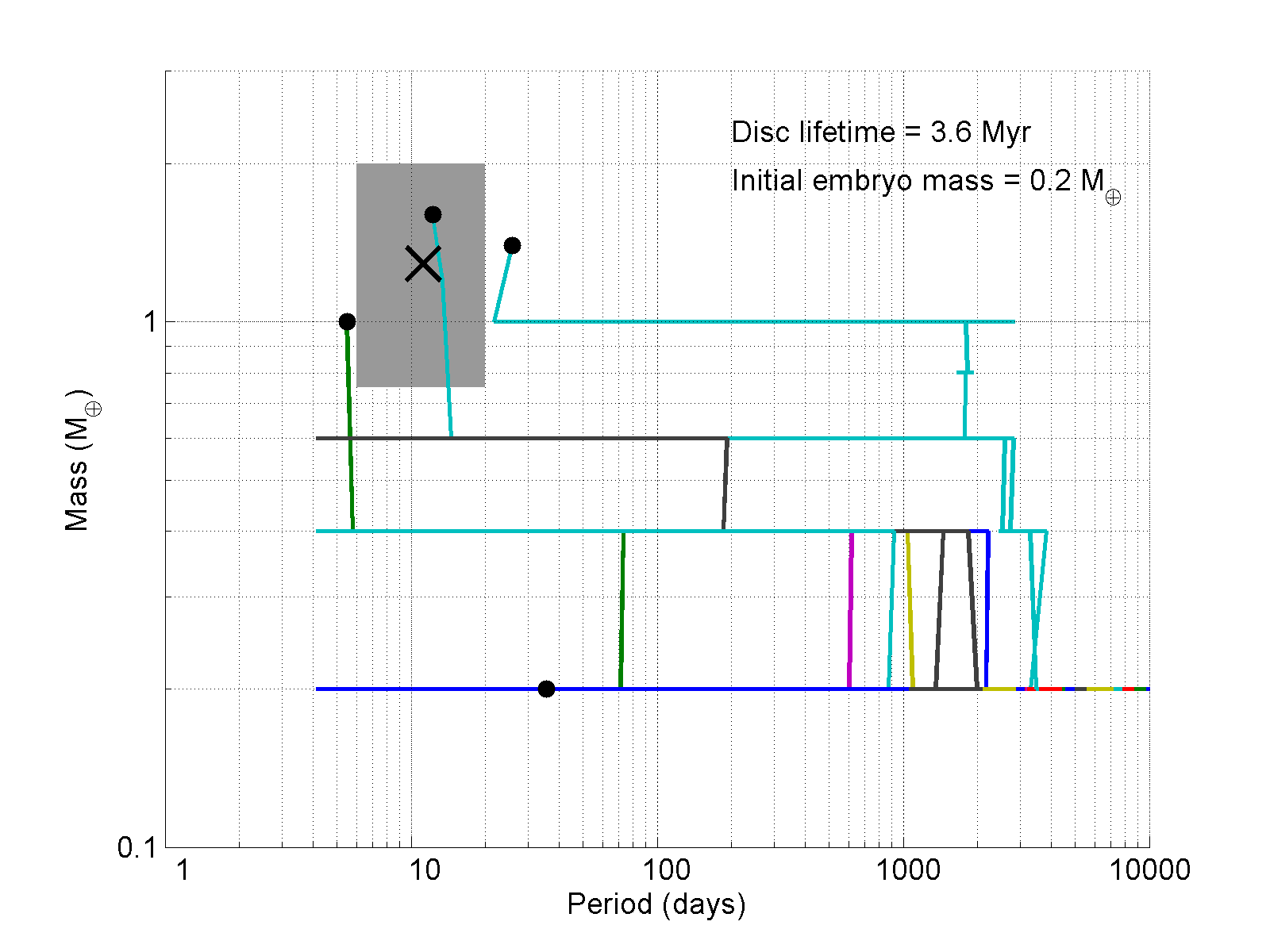}
\includegraphics[scale=0.42]{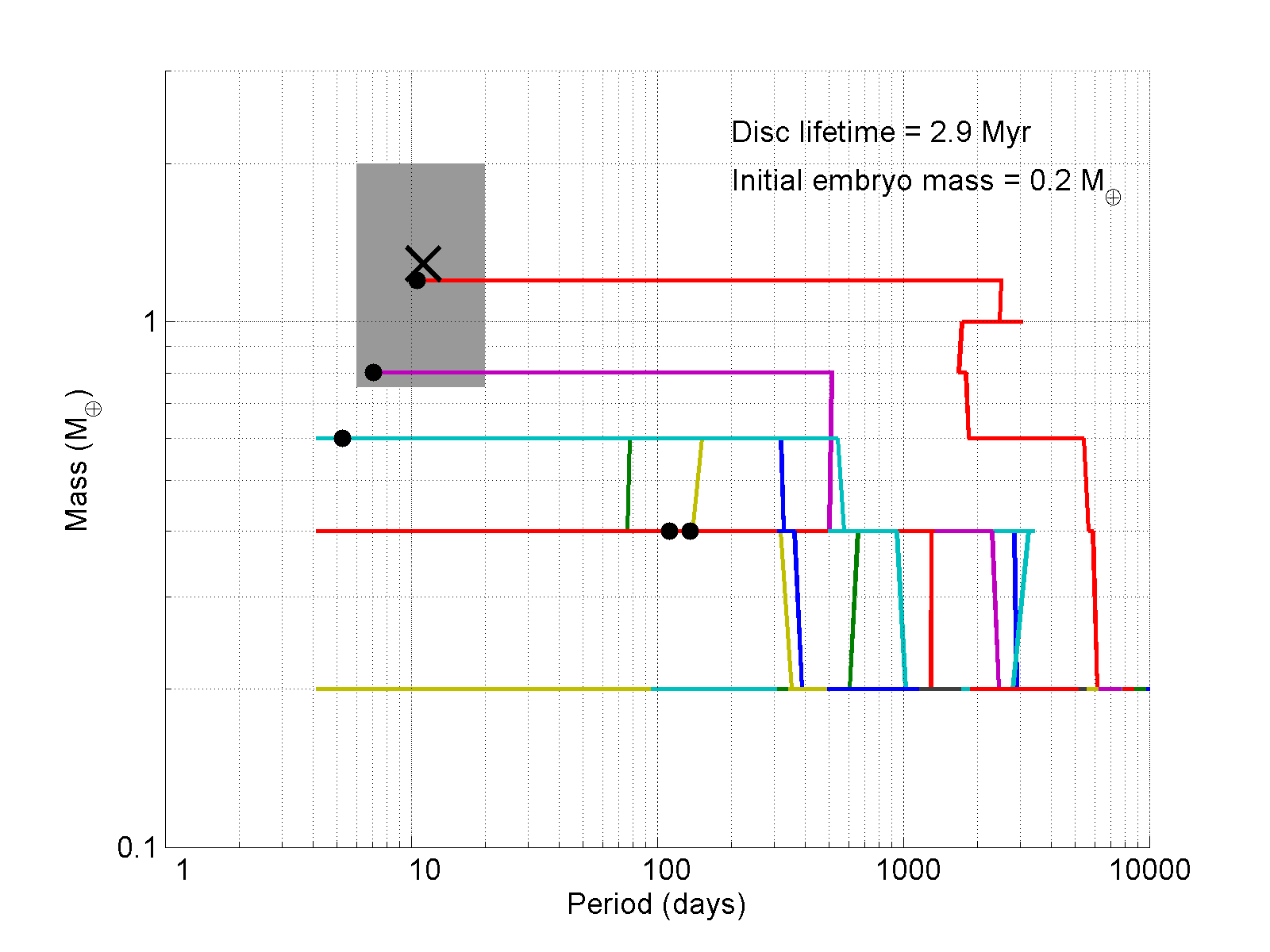}
\includegraphics[scale=0.42]{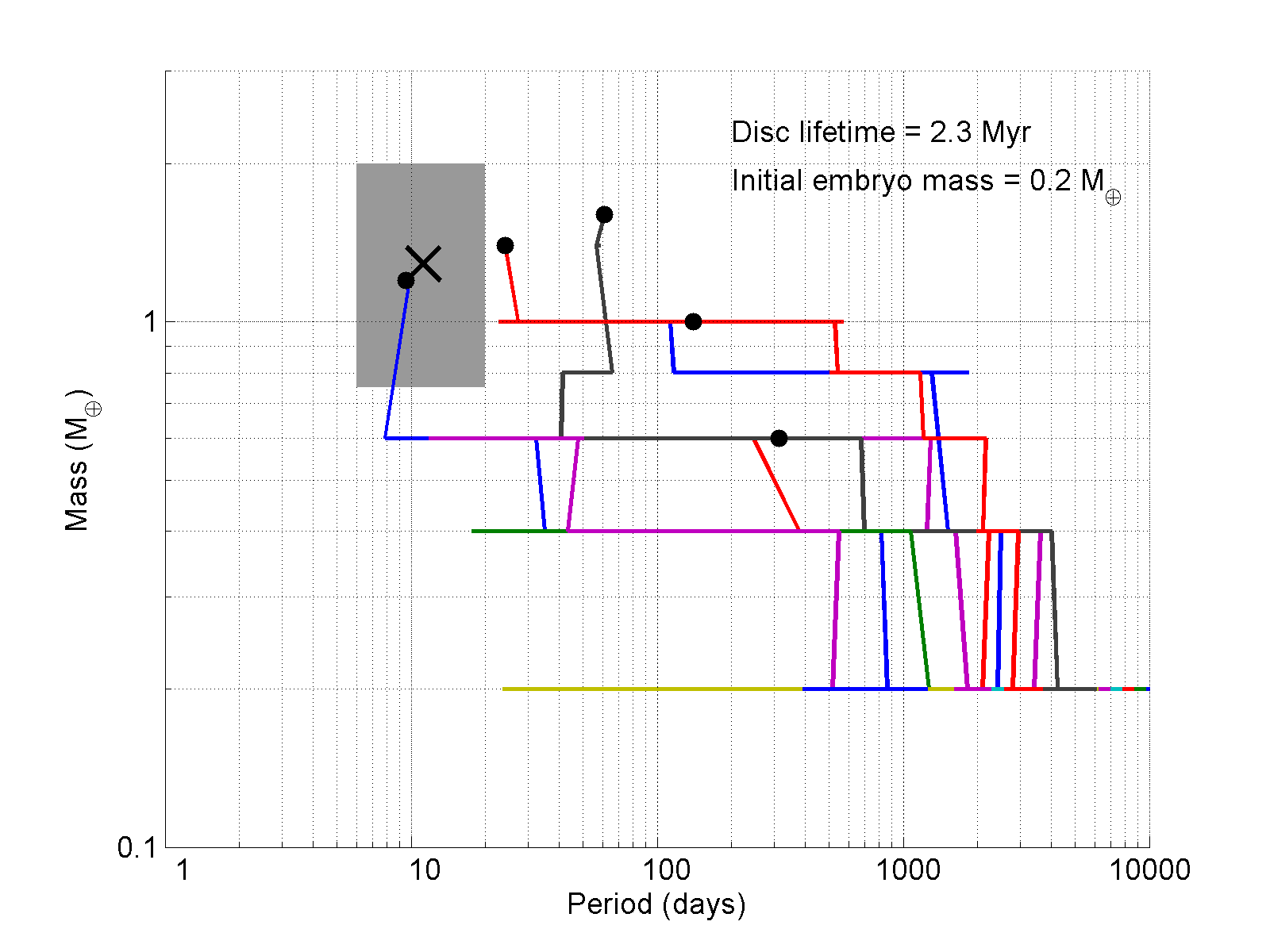}
\caption{Mass versus period evolution for four simulations in the multiple embryo formation scenario. The shaded grey region represents the mass and period range for planets that are classified as Proxima b analogues, and the black cross represents Proxima b.}
\label{fig:multipleembryo}
\end{figure*}

\subsection{Concurrent accretion and migration}
Instead of forming in situ, it could also have been possible for Proxima b to have formed at a much larger orbital radius, and then undergone migration to it's observed location, either as a single monolithic body, or as a collection of lower mass embryos that subsequently accreted through mutual collisions.
By forming further out in the disc beyond the snowline, it is highly likely that the planet would be composed of a significant abundance of water and other volatiles.
We examine two different formation scenarios where embryos form outside the snowline and migrate close to the star.
One scenario follows the evolution of multiple embryos as they migrate through a gas disc before reaching the inner regions of the system, where they can merge into one or more planets through undamped gravitational interactions at the end of the gas disc lifetime.
The second scenario focuses on how a single embryo forms at large orbital radius and migrates through a swarm of planetesimals, accreting a number of them, before halting its migration at an orbital period comparable to Proxima b at the end of the gas disc lifetime.
In both scenarios, we use the viscous disc models of \citet{ColemanNelson16}, adapted to include the initial disc parameters given in section \ref{sec:ProximaDisc}.

\subsubsection{Multiple migrating embryos}
\label{sec:multiple}
In the case of multiple embryos forming at large orbital radii, we assume that half of the dust in the disc remains small and contributes to the self-consistent calculation of the disc's opacity. In each of the simulations, some fraction of the remaining solids is assumed to have formed planetary embryos, with the remainder being lost from the system (implicitly assumed to be through accretion onto the star because of aerodynamic drag induced migration). These planetary embryos have an initial mass of 0.05, 0.1, 0.15 or 0.2 $\me$, initial semi-major axes between 1 and 9 $\au$, and are separated by 5 mutual Hill radii. The total mass in solids between different simulations ranges between 2.25 and 6 $\me$, which results in a conversion efficiency from dust to planetary embryos of between 10 and 25 \% when compared with the total dust abundance in the Proxima disc.
The embryos are able to accrete other embryos and can migrate within the viscously evolving gas disc, which we take to be the Proxima disc discussed in section \ref{sec:ProximaDisc}. The disc dissipates through photoevaporation after between 1.6 and 4.2 Myr, depending on the adopted photoevaporation rate.
We run the simulations for a runtime of 100 Myr to allow the planetary systems to continue evolving through scattering and collisions after the dispersal of the Proxima disc.

Embryos embedded within a gas disc have specific migration behaviour determined by their mass and orbital position.
We find that these migration behaviours and subsequent `migration maps' are similar to those displayed in \cite{ColemanNelson16}, except the regions of outward migration are pushed down to lower masses and also occur closer to the central star, as shown in fig. \ref{fig:contours}.
As the embryos migrate, there can be numerous close encounters, with some leading to collisions that create more massive embryos.
As an embryo's mass increases, it begins to migrate inwards at a faster rate, forcing other slower migrating embryos to migrate more quickly as part of a resonant convoy.
Over the course of the disc lifetime, the embryos continue to migrate into the inner regions of the system, with occasional collisions between embryos resulting in mass growth.

When the gas disc comes near to the end of its lifetime, most of the surviving embryos have orbital periods less than 1000 days, and masses $m_{\rm p} \le 1 \me$.
Embryos that have not survived have either been accreted by other more massive embryos, or have migrated past the inner edge of the disc and on to Proxima.
For the surviving embryos, a period of chaotic orbital evolution and collisions can often follow, as a result of embryo eccentricities not being damped due to the removal of the local gas disc.
These collisions lead to only a handful of embryos surviving after 100 Myr. In cases where most of the planetary growth occurred during the gas disc lifetime, the final systems often had planets occupying mean-motion resonances.

Figure \ref{fig:multipleembryo} shows the mass versus period evolution for a selection of simulations, with the black dots representing the final planet masses and orbital periods.
All of these simulations produced a planet that has similar characteristics to Proxima b, i.e. in terms of its mass and orbital period.
These planets formed from embryos that were initially located at large orbital radii, beyond the snowline, before migrating into the inner regions of the system, typically as part of a resonant convoy with other embryos.
When the resonant convoys reached the inner regions of  the system, the gas disc was near to the end of its lifetime, halting migration of the resonant convoys at short orbital periods.
With a lack of gas disc damping, embryo eccentricities were able to rise, resulting in collisions between neighbouring embryos.
After a chaotic period, the embryos were able to enter into a stable orbital configuration leaving the systems in the final states shown in Fig.~\ref{fig:multipleembryo}.

\begin{figure}
\includegraphics[scale=0.45]{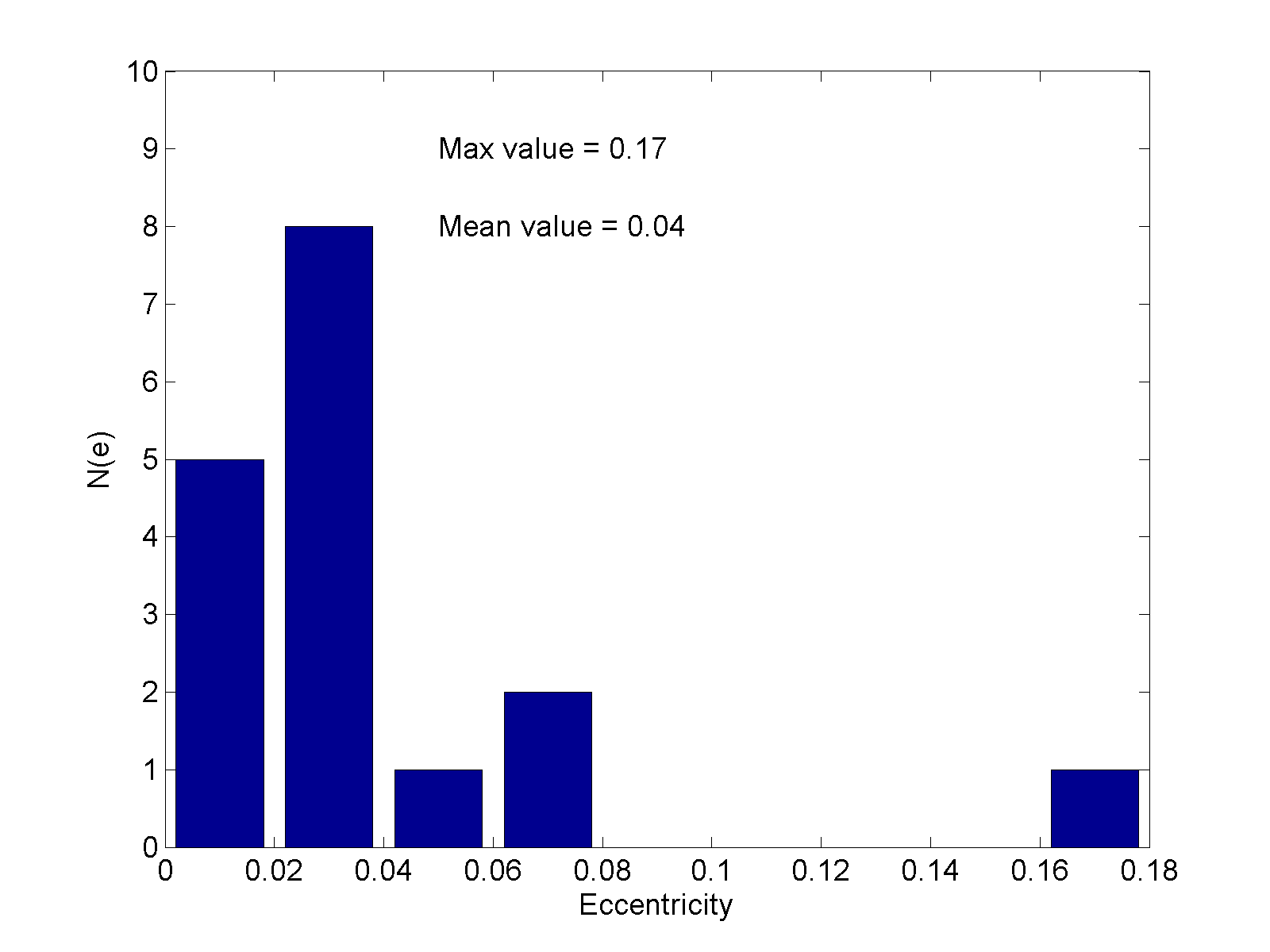}
\caption{Eccentricity distribution of Proxima b analogues formed in the multiple migrating embryo formation scenario. We define a Proxima b analogue as a planet with mass $0.75 \leq m_{\rm p} \leq 2\me$, and orbital period $6\leq P \leq20$ days.}
\label{fig:mmecc}
\end{figure}

\begin{figure}
\includegraphics[scale=0.45]{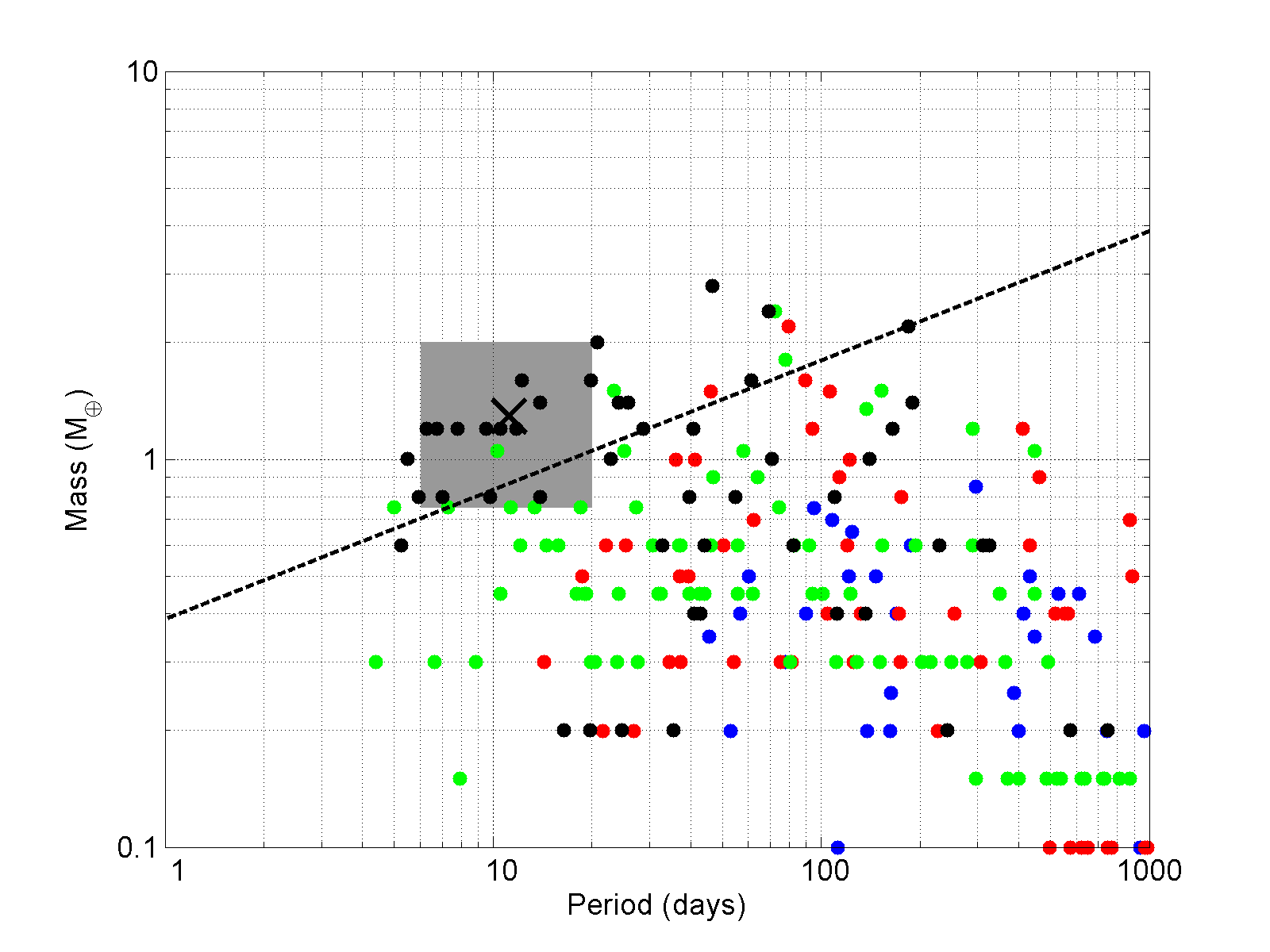}
\caption{A mass versus period plot for all surviving planets in the multiple migrating embryo formation scenario. Different colours represent the initial total mass of solids: blue - 2.25 $\me$, red - 3.6 $\me$, green - 4.65 $\me$, black - 5.8 $\me$. The shaded grey region represents the mass and period range for planets that are classified as Proxima b analogues, and the black cross represents Proxima b.
The dashed line represents a radial velocity detection threshold of K$\sim1$ m/s.}
\label{fig:mvpmm}
\end{figure}

These final systems typically contain a number of planets, and though this is not surprising since initially there were numerous planetary embryos, it is significant that these embryos do not form one single planet.
Instead, they form one or two planets that contain a significant bulk of the mass, accompanied by a small number of less massive planets.
Since the Proxima b analogues are accompanied by other planets, they will also have non-zero eccentricities.
Figure \ref{fig:mmecc} shows the eccentricity distribution of all Proxima b analogues, where the average eccentricity is $e \sim 0.04$, slightly reduced to that found in the in situ formation scenario.
In both scenarios it is the interactions and collisions between planetary embryos/planetesimals in a gas-free environment that increases eccentricities.
Figure \ref{fig:mvpmm} shows the final mass versus period distribution of all planets formed in this scenario.
It highlights that there are a significant number of Proxima b analogues, as well as a large number of less massive planets, as well as a number of more massive planets with longer periods, that would be difficult to detect in current radial velocity surveys, where the dashed line indicates a radial velocity amplitude of 1 m/s and so only objects lying above that line could be detected using current observations.

When examining the planetary compositions, we find that since the planets formed and accreted a significant amount of material outside the snowline, they have high volatile abundances. Although photolysis and the escape of hydrogen are likely to significantly alter the composition of the atmosphere, the large inventory of water carried by the Proxima b analogues means that they should retain large abundances of water. Indeed, they should be considered as `Ocean planets'.

\subsubsection{A single migrating and accreting core}
\label{sec:singlecore}
For this scenario we assume that a single planetary embryo has been able to form at a large orbital radius, and that half of the remaining solids in the disc have grown efficiently from dust into planetesimals (with the remaining dust providing opacity in the disc).
In each simulation, planetary embryos have an initial mass of 0.05, 0.1 or 0.2 $\me$, and are assigned initial semi-major axes between 1--9 $\au$.
Planetesimals are distributed between 0.5 and 9.5 $\au$, have a mass 20 times smaller than the planetary embryos, and have physical radii that are 10m, 100, 1km or 10km (a single size is adopted in each run). The gas disc is the Proxima disc model discussed in Section \ref{sec:ProximaDisc}, with an outer disc radius of 10$\au$ and a lifetime of 2.9 Myr.

As planetary embryos migrate through the gas disc, they encounter planetesimals, and accrete a number of them.
The increase in mass of the planetary embryos allows them to migrate inwards at a faster rate due to increased Lindblad torques.
This increased migration continues until the embryos reach a zero migration zone, where corotation torques arising from gas orbiting in an embryo's co-orbital region balance the Lindblad torques acting on the embryo.
The zero migration zones then drift inwards as the gas disc cools and dissipates, allowing the embryos to drift in with them on the disc evolution time \citep{Lyra10}, accreting planetesimals from their local vicinity.
The inward migration of embryos continues until the gas disc fully dissipates or the embryos accrete enough planetesimals for their corotation torques to saturate because of the increased mass of the planet, allowing the embryos to rapidly migrate inwards and on to the central star.

\begin{figure}
\includegraphics[scale=0.45]{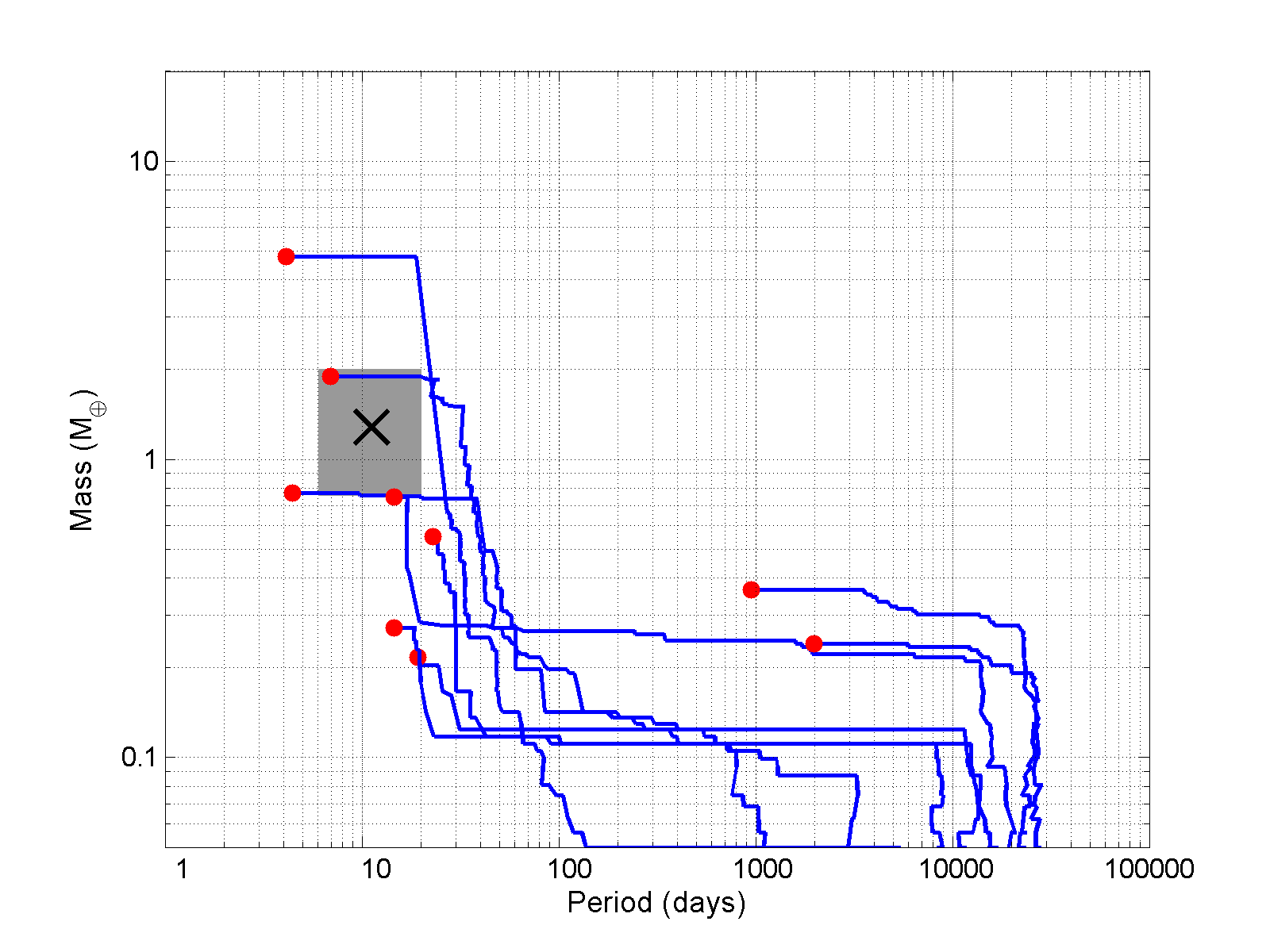}
\caption{Evolution tracks for single planetary embryos of mass 0.05$\me$ embedded in a gas disc with planetesimals of radius 100m. The red dots represent the embryos final masses and periods. The shaded grey region represents the mass and period range for planets that are classified as Proxima b analogues, and the black cross indicates the values for Proxima b.}
\label{fig:singleembryo}
\end{figure}

Figure \ref{fig:singleembryo} shows the evolution tracks of nine embryos, with initial masses $m_{\rm p} = 0.05\me$, migrating in a planetesimal disc where the planetesimals are 100m in size.
The evolution of the most massive object found in fig. \ref{fig:singleembryo} closely follows the description above of forming a Proxima b analogue.
This planet accreted a significant number of planetesimals such that it was then able to migrate inwards to a zero migration zone.
Once at the zone, the planet's migration was halted due the balance between corotation and Lindblad torques.
The planet then migrated with the zero migration zone into the inner disc, accreting numerous planetesimals before finishing its migration as a Proxima b analogue.

\begin{figure}
\includegraphics[scale=0.45]{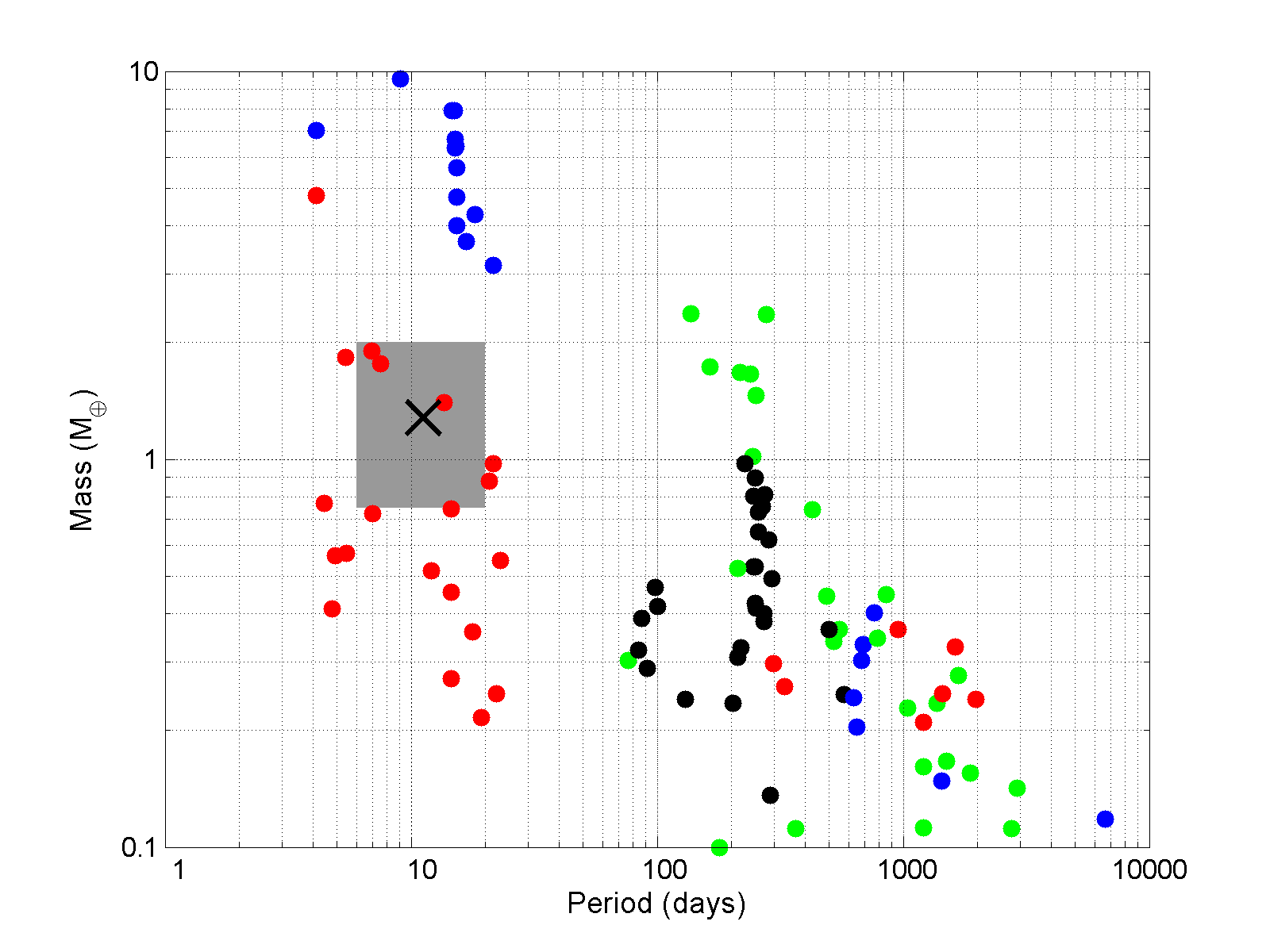}
\caption{A final mass versus period plot for all simulated planets in the single embryo formation scenario. Different colours represent different planetesimal sizes: blue - 10m, red - 100m, green - 1km, black - 10km. The shaded grey region represents the mass and period range for which planets are classified as Proxima b analogues, and the black cross indicates the values for Proxima b.}
\label{fig:singleembryooverall}
\end{figure}

Figure \ref{fig:singleembryooverall} shows the final masses and periods of all planetary embryos in these simulations.
As can be seen, there is a diversity in the final masses and periods of the accreting embryos.
Only a few embryos here have been able to obtain a mass and orbital period similar to Proxima b, with a few others coming close to being a Proxima b analogue, but these did not migrate inwards sufficiently or accrete enough planetesimals.
A few other planets accreted too many planetesimals late in the disc lifetime allowing them to become super-Earths, whilst the majority of the other planets were not able to accrete significant amounts of planetesimals, and as such underwent significantly less migration leaving them on long orbital periods.
This difference in accretion and migration rates is primarily due to the size of planetesimals, with smaller boulders allowing large accretion and migration rates, and larger planetesimals yielding much smaller migration and accretion rates \citep{ColemanNelson16}.
Since the majority of the mass growth that occurred for the Proxima b analogue did so outside the snowline, this means that this planet has a high quantity of volatile material including water, amounting to $\sim 66\%$ by mass \citep{Lodders2003}.

\subsection{Consolidation of pebbles}
\label{sec:pebbles}
We now consider the formation of Proxima b through the accretion of pebbles onto a planetary embryo.
Here we implement the pebble accretion model of \citet{Lambrechts14} within the gas disc model of \citet{ColemanNelson16}.
We assume that a single large accreting planetesimal has been able to form in the disc, and is able to migrate and accrete pebbles.
Planetesimals are given initial semi-major axes of between 0.1 and 9 $\au$ to examine pebble accretion in different regions of the disc, and they have initial masses equal to the local pebble transition mass, the mass at which efficient pebble accretion begins to operate.
We again use the Proxima disc model discussed in Section \ref{sec:ProximaDisc} with an outer radius of 10$\au$ and a lifetime of 3 Myr.

\begin{figure*}
\includegraphics[scale=0.42]{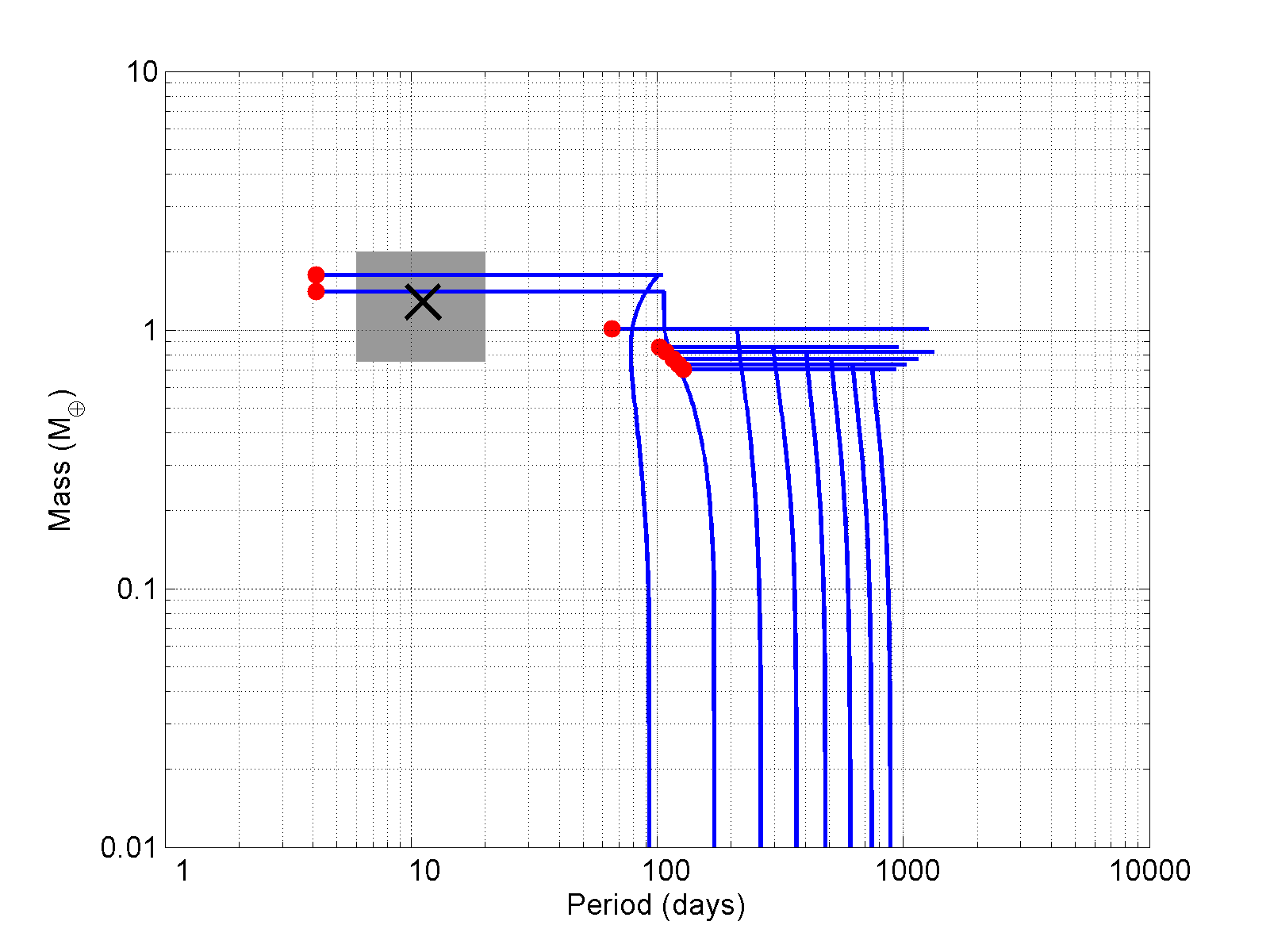}
\includegraphics[scale=0.42]{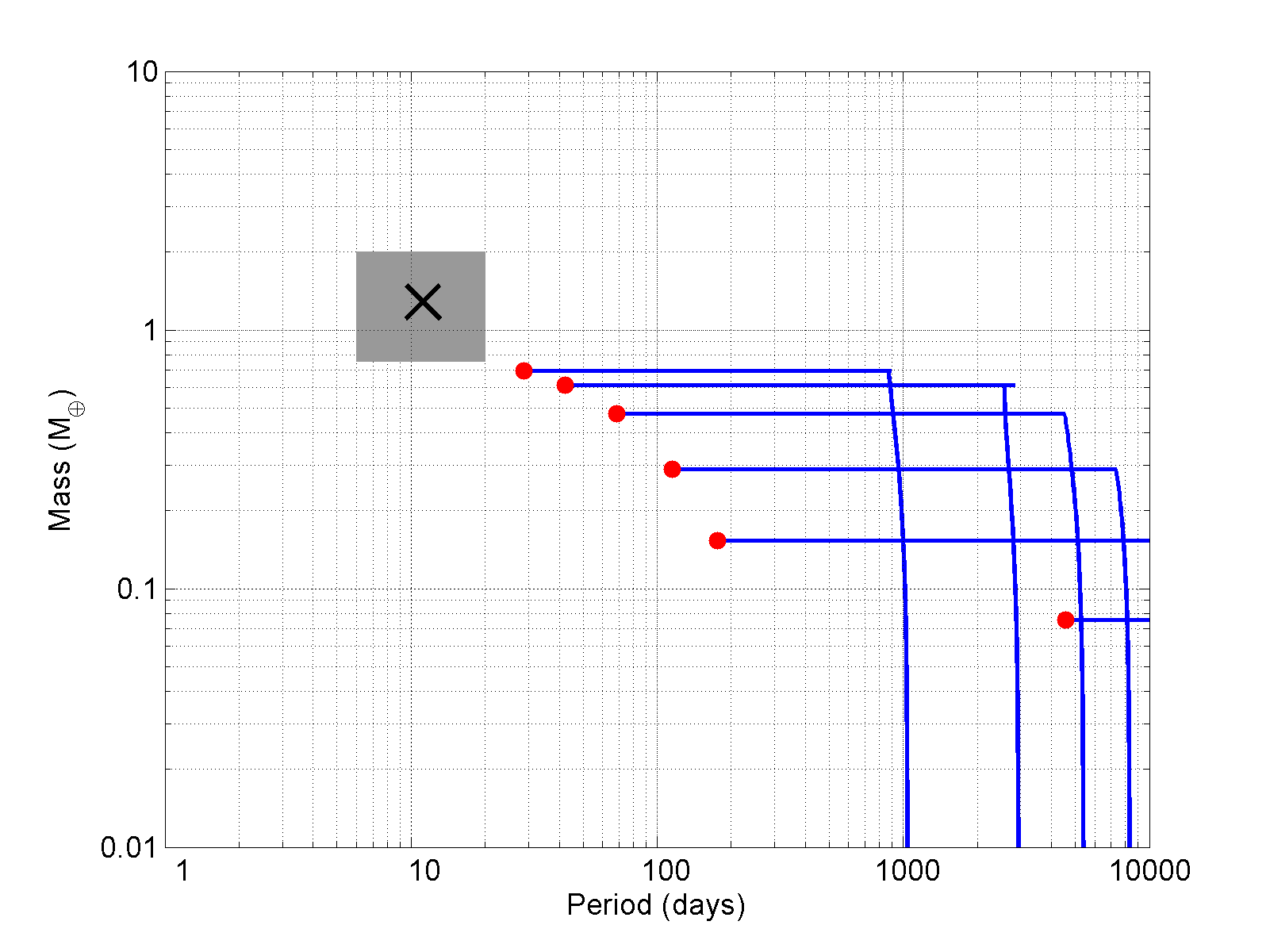}
\caption{Evolution tracks for single planetary embryos accreting dry (left panel) and icy (right panel) pebbles in a Proxima disc with a lifetime of 2.9 Myr. The red dots represent the embryos' final masses and periods. The shaded grey region represents the mass and period range for which planets can be classified as Proxima b analogues, and the black cross represents Proxima b.}
\label{fig:pebbles}
\end{figure*}

As the protoplanetary disc evolves, a pebble formation front moves outwards from the centre of the system, and the forming pebbles migrate inwards through the disc due to aerodynamic drag.
When the pebble growth front reaches and extends out beyond a planetesimal's orbital radius, if the planetesimals mass is above a `transition mass', then it is able to accrete pebbles efficiently as they drift past the planetesimal. Mass growth of the planetesimal can be fast here depending on the abundance of pebbles drifting past the planetesimal, as well as the gas drag forces acting on the pebbles, increasing the planetesimal's capture radius \citep{Lambrechts12,OrmelKlahr2010}.
Accretion of pebbles continues until either the creation of pebbles ceases as the pebble front reaches the outer edge of the disc, or when planetesimals reach their isolation masses. At this point the accreting planetesimals have grown to become planets, and the planet's gravity perturbs the gas disc, halting pebble accretion through the creation of a small pressure bump \citep{MorbidelliNesvorny12}.
After pebble accretion is halted, the planet undergoes migration, until it either migrates into the central star, or the gas disc fully dissipates. In our models, we consider that in each simulation a single accreting planetesimal can form within a range of radii, and we then track the mass growth and migration of this object as it grows to become a planet.

Figure \ref{fig:pebbles} shows the evolution tracks for embryos accreting dry (left panel) and icy (right panel) pebbles, by virtue of being located initially inside or outside of the snowline, respectively.
{As pebbles drift through the snowline into the inner regions of the disc, their water and other volatile components sublimate, leaving only rocky pebble cores available for accretion.
In both panels, we see that embryos accrete pebbles efficiently until the supply of pebbles diminishes as the pebble production front reaches the outer edge of the disc.
Since embryos at larger orbital periods begin to accrete pebbles at a later time, they have less time overall to accrete pebbles and as such have smaller final masses.
Once the embryos reach a mass $m_{\rm p}>0.1\me$ they begin to significantly migrate in the disc, migrating to zero-migration zones where corotation and Lindblad torques balance each other.
Embryos migrate with the zero-migration zones as they slowly drift inwards as the disc cools and dissipates.
The embryos continue to slowly migrate inwards until the disc fully dissipates.

When considering the embryos that accrete dry pebbles, we can produce a planet that has a similar mass and period to Proxima b. Although our models do not produce a Proxima b analogue as we have defined it, we note that moderate changes to the underlying disc model that we consider would result in such an object forming.
However, it becomes more difficult to form a Proxima b analogue when considering embryos that accrete water-rich pebbles outside the snowline.
This is due to the lack of available pebbles for accretion, either because the mass of solids in the disc is insufficient, or because the pebble production front reaches the outer edge of the disc before significant mass growth has occurred.
For a planet to form that is a true Proxima b analogue, a greater amount of solids is required, i.e. a larger disc mass would be required for fixed disc metallicities, which could be achieved by increasing either the surface density or by extending the disc radius.

Follow up observations that are able to characterise Proxima b's water and volatile content would be able to narrow down the location that the planet formed if we assume that it underwent pebble accretion, particularly if the planet turned out to be a completely dry world as this would indicate formation interior to the snowline in the disc.

\section{Dynamical stability of Proxima~b's orbit in the presence of an additional super-Earth}
\label{sec:dynamical}
The interpretation of the Doppler data for Proxima in the period range between 50 and 500 days remains ambiguous.
While activity is likely responsible in part for the observed variability, the variability would also be consistent with the presence of additional planets in the super-Earth mass regime.
The presence of additional companions might be very relevant to the putative properties of planets like Proxima b, as additional planets tend to produce non-zero eccentricities thus affecting the possible rotation states and tides.
The dynamical evolution of such planetary systems was investigated using the Mercury-6 code \citep{Chambers}.
Following \citet{Bolmont13} we implemented tidal forces using a constant time lag model assuming pseudo-synchronous rotation.
We also implemented the first order post-Newtonian correction in Mercury-6.

We simulated the Proxima system with Proxima b according to the parameters determined from the radial velocity variations.
We added a potential second planet in order to check the dynamical stability.
The dynamical evolution has been followed over at least 10$^7$ years to prove secular stability of the system.
We used the Mixed Variable Symplectic (MVS) integrator implemented in Mercury-6 with an integration step of 0.5 days (i.e. less than a twentieth of the orbital period of the inner planet).
Tidal forces are taken into account only for the inner planet.
The tidal dissipation as well as the Love number of degree 2 were chosen to be compatible with Earth's.
The planetary radius of Proxima b was estimated from the minimum mass, assuming a mean density equal to that of the Earth.
We assume that Proxima b starts in a state of synchronous rotation, and we assume 83 days as the rotation period of Proxima.

Since the parameter range with a rather unconstrained outer planet is huge, we selected a {\it stress test} model.
Here Proxima b has the highest possible eccentricity (e=0.35).
For the putative Proxima c, we assume an eccentricity of 0.3 and a mass of 10 Earth masses.
The largest effect is expected, when the two planets are in mean motion resonance, so we placed Proxima c at the low end of the possible period range in a 5:1 resonance with Proxima b ($\sim 56$ days).
The orientation of the pericenters, as well as the starting configuration, were chosen to have the two planets in the closest possible configuration (i.e. Proxima b in apastron, Proxima c in periastron).
The system survives for at least 10$^7$ years with, however, a significant interaction as is shown in fig. \ref{fig:dynamicalStudy}.
Since the {\it stress test} model is unfavourable for the stability of the system, this can be regarded as a worst case scenario, but shows that dynamical simulations can be used to place constraints on the parameter space for additional planets in the system.
With a lower mass of Proxima c, and/or a longer orbital period, the interactions are significantly reduced, such that the system would be stable for much longer than the executed integration times ($\gg 10^7$ years) of the stress test.

\begin{figure}
\center
\includegraphics[scale=0.42]{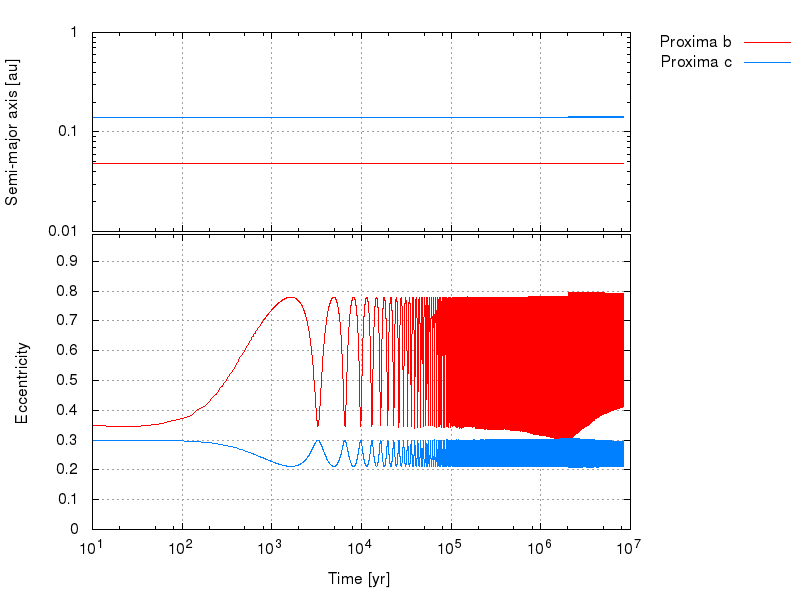}
\caption{Dynamical evolution of the planetary system integrated by Mercury-6. Proxima b is set to the maximum eccentricity consistent with the Doppler data and the potential Proxima c with assumed mass of 10 Earth masses is at the lower limit of the period range possible. As a {\it stress test} it is in 5:1 resonance with b and the orientation of the ellipses minimize the smallest distance between the two.}
\label{fig:dynamicalStudy}
\end{figure}

\section{Discussion and conclusion}
\label{sec:conc}
We have undertaken a study of four scenarios for the formation of the planet Proxima b to examine how they differ in terms of their predictions about planetary multiplicity, volatile content and orbital eccentricity that may be tested by future observations.

In scenario (i), we examined the types of planetary systems that arise when the formation of Proxima b analogues arises through mutual impacts between planetary embryos and planetesimals in situ after the dispersal of the gas disc. In agreement with previous studies in application to both the Solar System and extrasolar planet systems \citep{ChambersWetherill1998, HansenMurray2012}, we find that this formation scenario always results in the formation of multiple planet systems. In those cases where a decent Proxima b analogue was formed in our runs, we obtained systems where the additional planets would be below current radial velocity detection thresholds of 1 m/s, and systems in which multiple detectable planets were formed. Although our initial disc of planetary embryos and planetesimals did not extend out as far as the snowline, we examined the sizes of the feeding zones for the Proxima b analogues that formed, and found that they always encompassed the outer edge of our initial population of planetesimals. This suggests that the dynamical diffusion of planetesimals is quite efficient, as has been found by previous studies \citep[e.g.][]{Raymond2007}, indicating that a disc of embryos and planetesimals that extends out beyond the snowline might allow a moderate flux of icy planetesimals to accrete onto the Proxima b analogues, delivering an inventory of water and other volatiles. As such, it remains a distinct possibility that formation via this scenario could result in Proxima b being endowed with volatiles, including water, sufficient to provide a secondary atmosphere and an ocean. Examination of the final orbital eccentricities of the Proxima b analogues arising in the simulations indicates that they cover a range of values between 0.02 -- 0.23. Small values of the eccentricity ($e \le 0.05$) would likely result in Proxima b becoming tidally locked to its star in a  1:1 spin-orbit resonance, but larger values could lead to Proxima b becoming caught in a 3:2 spin-orbit resonance, similar to Mercury in the Solar System, as discussed recently by \citet{Ribas16}. It is interesting to note that \citet{Ribas16} examine the tidal evolution of eccentricity, and show that for a rate of dissipation of tidal energy that is a factor of 10 smaller than on Earth, perhaps appropriate for an `Ocean world', and also for a larger value that is appropriate to the Earth, the eccentricity damping time is longer than the age of the Proxima system (4.8 Gyr). As such, future observations should be able to better constrain the eccentricity of Proxima b and shed some light on its formation history. 

In scenario (ii), we examined the outcome of accretion between a set of multiple embryos whose orbital radii extend across a large range of distances from the star. The embryos are embedded in a gaseous protoplanetary disc that drives orbital migration. Not surprisingly, this scenario also always results in systems of multiple planets, and the Proxima b analogues are often found to be in mean motion resonances, similar to those produced in previous work \citep{CresswellNelson2006, TerquemPapaloizou2007, Cossou14, ColemanNelson16}. Given that most of the embryos originated outside of the snowline, the planets that form in this scenario have water contents that are $\sim 66\%$ by mass, although we note that this is likely to be an overestimate due to the fact that the simple hit-and-stick accretion prescription that our simulations adopt neglects devolatisation. Nonetheless, it seems likely that bodies formed according to this scenario will be `Ocean worlds' that do not have significant land masses. It is interesting to note that Proxima b was orbiting interior to the inner edge of the classical habitable zone during the first $\sim 200$ Myr of Proxima Centauri's life time due to evolution of the star's luminosity, such that water on the planet may have been in vapour phase during this time. The calculations of \citet{Ribas16} and \citet{Barnes2016}, however, suggest that only a modest fraction of the planet's water would be lost during this time through photolysis and the loss of hydrogen, such that the planet would still possess a substantial inventory of water once the habitable zone moves inwards and encompasses the planet's orbit. Although the eccentricities of the Proxima b analogues formed at the end of the disc lifetime in this scenario tend to be smaller than those formed in scenario (i), dynamical interactions within the multiple planet systems after the gas disc has dissipated cause the eccentricities to grow to moderate values, and in some cases large enough to allow the capture of Proxima b into a 3:2 spin-orbit resonance.

In scenario (iii), we examined the formation of Proxima b analogues through the accretion of planetesimals onto a single migrating embryo. By construction, this model results in a single water-rich `Ocean world' on a circular orbit. 

Finally, in scenario (iv) we examined the formation of Proxima b through pebble accretion onto a single embryo that can form within a broad range of orbital radii. We found it more difficult to form a viable Proxima b analogue within this scenario due to our fixed assumptions about the underlying disc model, but with modest relaxation of these assumptions a Proxima b analogue could be formed easily. As with scenario (iii), this produces a single planet on a circular orbit. We find cases where the planet can be a dry world because pebble accretion occurs exclusively interior to the snowline, and cases where water-rich planets are formed by pebble accretion outside the snowline.
Though here we have only examined pebble accretion onto a single embryo, it is worth noting that multiple embryos could undergo pebble accretion \citep{Levison15b,Chambers16}. If this were the case, we would expect the outcomes to be similar to that of scenario (ii), but still with the possibility of dry worlds forming interior to the snowline.

We also examined the dynamical stability of the system if an additional super-Earth mass planet was located in 5:1 resonance with Proxima b.
We found that the planetary system in this \emph{stress test} was able to survive for at least $10^7$ years, however there was a significant interaction, slightly altering both planet's semi-major axes and eccentricities.
Though this \emph{stress test} is considered unfavourable for a planetary system's stability, it essentially constitutes a worst case scenario, and it shows that dynamical simulations can be used to constrain the parameter space for any additional planets in the Proxima system.
Considering the additional planet had the highest possible mass, and was situated in one of the lowest possible orbital periods consistent with the data, we expect that with a lower mass or longer period for the additional planet, the interactions between the two planets would be significantly reduced, eventually lasting for much longer than the estimated age of the Proxima system ($\sim4.8$ Gyr).

In summary, follow up observations of Proxima b that indicate that it is a member of a multiple system that consists of planets with moderate water-contents would favour an origin similar to scenario (i) above. If Proxima b is found to be a member of a multiple system of Ocean planets then scenario (ii) would be favoured.
If Proxima b is found to be on a moderately eccentric orbit, then scenarios (i) or (ii) would be viable, whilst if Proxima b had a circular orbit, then scenarios (iii) or (iv) would be favoured. If Proxima b turns out to water-rich and the only short-period planet orbiting Proxima Centauri, then either scenarios (iii) or (iv) could provide a viable formation history. A single dry world would favour an origin through pebble accretion that arises interior to the snowline, assuming that significant water loss did not arise early in the planet's history. We await further observations that will be able to test some of these basic model predictions.

\section*{Acknowledgements}
The simulations presented in this paper utilised Queen Mary's MidPlus computational facilities, supported by QMUL Research-IT and funded by EPSRC grant EP/K000128/1.

\bibliographystyle{mnras}
\bibliography{references}{}

\end{document}